\documentclass[proceedings]{JHEP3}
\usepackage{amssymb}
\usepackage{amsfonts}
\usepackage{amsmath}
\usepackage{epsfig,multicol}

\newbox\mybox

\newcommand\fverb{\setbox\mybox=\hbox\bgroup\verb}
\newcommand\fverbdo{\egroup\medskip\noindent\fbox{\unhbox\mybox}\ }
\newcommand\fverbit{\egroup\item[\fbox{\unhbox\mybox}]}
\conference{Non-Hermitian systems from complex root spaces}
\abstract{We provide a general construction procedure for antilinearly invariant complex root spaces. The proposed method is generic and may be applied to any Weyl group allowing to take any element of the group as a starting point for the construction. Worked out examples for several specific Weyl groups are presented, focusing especially on those cases for which no solutions were found previously. When applied in the defining relations of models based on root systems this usually leads to non-Hermitian models, which are nonetheless physically viable in a self-consistent sense as they are antilinearly invariant by construction. We discuss new types of Calogero models based on these complex roots. In addition we propose an alternative construction leading to q-deformed roots. We employ the latter type of roots to formulate a new version of affine Toda field theories based on non-simply laced roots systems. These models exhibit on the classical level a strong-weak duality in the coupling constant equivalent to a Lie algebraic duality, which is known for the quantum version of the undeformed case.}

\title{Non-Hermitian multi-particle systems from complex root spaces}
\author{Andreas Fring and Monique Smith \\
Centre for Mathematical Science, City University London,\\
Northampton Square, London EC1V 0HB, UK\\
E-mail: a.fring@city.ac.uk , abbc991@city.ac.uk}

\input{tcilatex}

\begin{document}

\section{Introduction}

More than fifty years ago Wigner \cite{EW} observed that operators left
invariant under antilinear involutory transformations have real eigenvalues
when in addition their eigenfunctions also possess this symmetry. Based on
this fact one can regard such type of operators as being related to physical
observables in classical, quantum mechanical and even quantum field
theories. More recently \cite{Bender:1998ke,Benderrev} this feature was
exploited by taking $\mathcal{PT}$-symmetry, that is a simultaneous parity
transformation $\mathcal{P}$ and time reversal $\mathcal{T}$, as a concrete
realisation of this antilinear involutory map. Especially when the operator
is a quantum mechanical single particle Hamiltonian the symmetry is easily
identified and many new physically meaningful and self-consistent models
have been constructed. Also properties of older models could be explained
more rigorously by exploiting it. In contrast, for multi-particle systems or
field theories the deformations and the corresponding symmetries are less
obvious and may involve complicated transformations in the configuration
space. Often the symmetry becomes only apparent after a suitable change of
variables or even a full separation of variables has been carried out \cite%
{Milos,FZ}. Many interesting and even integrable multiparticle systems such
as Calogero-Moser-Sutherland models \cite{OP2} and also field theories such
as Toda field theories \cite{Wilson,DIO} are formulated generically in terms
of root systems associated to Weyl or Coxeter groups. The dynamical
variables or fields are in the dual space with respect to some standard
inner product. Since these root spaces are naturally equipped with various
symmetries due to the fact that by construction they are left invariant
under the action of the entire Weyl group, it is by far easier and
systematic to identify the antilinear symmetries in the root spaces rather
than in the configuration space. Once identified they can be transformed to
the latter.

This general idea was recently explored in \cite{FZ,FringSmith,FringSmith2},
were some antilinear symmetric deformations were identified for several Weyl
groups and the consequences were studied for some applications to modified
Calogero models. It was shown that under the assumptions made in these
papers the deformations with the desired properties did not exist for
certain Weyl groups. One of the purposes of this manuscript is to fill this
gap and provide solutions for the missing cases together with an explanation
of why they do not exist based on the previous constructions.

The main steps of the construction proposed here is to select \emph{any}
element $\hat{\omega}$ $\in \mathcal{W}$ of order two of the Weyl group,
i.e.~$\hat{\omega}$ is an involution $\hat{\omega}^{2}=\mathbb{I}$. This
element is then identified as the analogue of the parity operator $\mathcal{P%
}$. Subsequently $\hat{\omega}$, together with the root space it acts on, is
deformed in an antilinear fashion. This means the root spaces have to be
complex. One may also start from several elements and consequently construct
deformations invariant under the same amount of different antilinear
symmetries. Imposing further constraints on the number of the symmetries and
the nature of the deformation, such as demanding it to be an isometry and
possessing certain limiting behaviour, allows to determine it. Requiring
maximal symmetry in all simple Weyl reflections is only possible for groups
of rank 2. The explicit solutions for this scenario can be found for $A_{2}$%
, $G_{2}$ in \cite{FZ} and $B_{2}$ in \cite{Assis:2009gt}. In \cite%
{FringSmith} two possibilities have been investigated, to have a symmetry
with respect to two $\mathcal{P}$-operators identified as two factors of the
Coxeter element and one symmetry being the longest element of the Weyl
group. For the specific example of the $E_{8}$-Weyl group the option of two
symmetries giving rise to modified Coxeter transformations of order less
than the standard Coxeter number was explored in \cite{FringSmith2}. It is
mainly the latter construction which we generalise here, although the
proposed procedure is completely generic.

We shall also propose a construction of complex root spaces based on $q$%
-deformations, which arose in the context of the study of the
renormalisation of affine Toda field theories based on non-simply laced
algebras \cite{q1,q2}.

Clearly when applying these complex roots to define multi-particle systems
of Calogero or Toda type models will be non-Hermitian. However, due to the
build in antilinear symmetric invariance, the models are strong potential
candidates for physically meaningful models.

Our manuscript is organised as follows: In section 2 we lay out the
procedure of how to construct complex root spaces equipped with the desired
property to be antilinearly invariant. We exemplify this procedure in
section 3 for spaces which possess two antilinear symmetries which are
subfactors of the factors of the factorised Coxeter element. We provide
solutions for several Weyl groups for which hitherto no solutions were found
and for which it was even shown that solutions based on other assumptions do
not exist. In section 4 we explore the possibility to take these two
symmetries to be entirely arbitrary. In section 5 we reverse the
construction and start with given deformations based on simple rotations in
the configuration space and compute some of the corresponding root spaces.
An alternative deformation method leading to q-deformed roots with no
obvious antilinear symmetry is proposed in section 6. In section 7 and 8 we
apply our constructions to propose new types of Calogero-Sutherland-Moser
models and Toda field theories, respectively. We investigate some of the
features of these new models. Our conclusions and an outlook to future
investigations are presented in section 9.

\section{Construction of antilinearly invariant complex rootspaces}

We explain here the framework for a construction of complex extended
antilinearly invariant root systems which we denote by $\tilde{\Delta}%
(\varepsilon )$. We present a generalisation of a method introduced and
employed in \cite{FringSmith,FringSmith2} with focus on obtaining solutions
for cases which could not be found previously. The procedure consists of
constructing two maps, which may be obtained in any order. In one step we
extend the representation space $\Delta $ of the standard roots $\alpha $
from $\mathbb{R}^{n}$ to $\mathbb{R}^{n}\oplus i \mathbb{R}^{n}$. This means
we are seeking a map 
\begin{equation}
\delta :~\Delta \rightarrow \tilde{\Delta}(\varepsilon ),\qquad \alpha
\mapsto \tilde{\alpha}=\theta _{\varepsilon }\alpha ,  \label{defo}
\end{equation}%
where $\alpha =\{\alpha _{1},\ldots ,\alpha _{\ell }\}$, $\ \Delta \subset 
\mathbb{R}^{n}$, $\tilde{\Delta}(\varepsilon )\subset $ $\mathbb{R}%
^{n}\oplus i \mathbb{R}^{n}$ and $n$ is greater or equal to the rank $\ell $
of the Weyl group $\mathcal{W}$. The complex deformation matrix $\theta
_{\varepsilon }$ introduced in (\ref{defo}) depends on the deformation
parameter $\varepsilon $ in such a way that $\lim_{\varepsilon \rightarrow
0}\theta _{\varepsilon }=\mathbb{I}$. The deformation is constructed to
facilitate the root space $\tilde{\Delta}$ with the crucial property for our
purposes, namely to guarantee that it is left invariant under an antilinear
involutory map%
\begin{equation}
\varpi :\tilde{\Delta}(\varepsilon )\rightarrow \tilde{\Delta}(\varepsilon
),\qquad \tilde{\alpha}\mapsto \omega \tilde{\alpha}.  \label{map}
\end{equation}%
This means the map in (\ref{map}) satisfies $\varpi :\tilde{\alpha}=\mu
_{1}\alpha _{1}+\mu _{2}\alpha _{2}\mapsto \mu _{1}^{\ast }\omega \alpha
_{1}+\mu _{2}^{\ast }\omega \alpha _{2}$ for $\mu _{1}$, $\mu _{2}\in 
\mathbb{C}$ and $\varpi \circ \varpi =\mathbb{I}$.

We assume next that $\omega $ decomposes into an element of the Weyl group $%
\hat{\omega}$ $\in \mathcal{W}$ with $\hat{\omega}^{2}=\mathbb{I}$ and a
complex conjugation $\tau $, $\omega =\tau \hat{\omega}=\hat{\omega}\tau $.
The presence of $\tau $ ensures the antilinearity of $\varpi $. In some
concrete applications it is understood that the maps $\hat{\omega}$ and $%
\tau $ correspond to analogues of the parity $\mathcal{P}$ and time reversal
operator $\mathcal{T}$, respectively. Candidates for $\hat{\omega}$
previously explored are simple Weyl reflections $\sigma _{i}$ \cite{FZ}, the
two factors $\sigma _{\pm }$ of the Coxeter element \cite{FringSmith}, the
longest element $w_{0}$ of the Weyl group \cite{FringSmith} and some more
general elements in $\mathcal{W}$ for the example of $E_{8}$ \cite%
{FringSmith2}. It is the latter construction which we focus on here and
extend in a more general and systematic way.

Concretely we assume here that we have at least two different involutions $%
\varpi $ of the type (\ref{map}) at our disposal, say $\varpi _{i}$ with $%
i=1,2,\ldots $ With our application in mind, namely to construct physically
viable self-consistent non-Hermitian multi-particle systems, one such map
would in principle be sufficient. However, the presence of two maps leads
immediately to some extremely useful constraints. We take the associated
rules of correspondence to be of the form%
\begin{equation}
\omega _{i}:=\theta _{\varepsilon }\hat{\omega}_{i}\theta _{\varepsilon
}^{-1}=\tau \hat{\omega}_{i},\qquad \text{for }i=1,\ldots ,\kappa \geq 2.
\label{o1}
\end{equation}%
Then by 
\begin{equation}
\omega _{i}\omega _{j}=\tau \hat{\omega}_{i}\tau \hat{\omega}_{j}=\tau ^{2}%
\hat{\omega}_{i}\hat{\omega}_{j}=\hat{\omega}_{i}\hat{\omega}_{j}=\theta
_{\varepsilon }\hat{\omega}_{i}\hat{\omega}_{j}\theta _{\varepsilon }^{-1},
\end{equation}%
it follows directly that the composition of any two of these elements of the
Weyl group $\Omega _{ij}:=\hat{\omega}_{i}\hat{\omega}_{j}$ commutes with
the deformation matrix $\theta _{\varepsilon }$%
\begin{equation}
\left[ \Omega _{ij},\theta _{\varepsilon }\right] =0.  \label{ohm}
\end{equation}%
Note that in general $\Omega _{ij}\neq \Omega _{ji}$. Examples previously
considered \cite{FringSmith} were for instance $\hat{\omega}_{1}=\sigma _{-}$
and $\hat{\omega}_{2}=\sigma _{+}$, such that $\Omega _{12}=\sigma $. Since
by construction $\Omega _{ij}\in \mathcal{W}$ we can expand $\theta
_{\varepsilon }$ in all elements $\check{\omega}_{i}\in \mathcal{W}$ which
commute with $\Omega _{ij}$, i.e.~$\left[ \Omega _{ij},\check{\omega}_{i}%
\right] =0$, 
\begin{equation}
\theta _{\varepsilon }=\sum\limits_{k}r_{k}(\varepsilon )\check{\omega}%
_{k}\qquad \text{for }r_{k}(\varepsilon )\in \mathbb{C},  \label{exp}
\end{equation}%
and subsequently determine the coefficient functions $r_{k}(\varepsilon )$
from additional constraints. One further natural constraint is to demand
that $\theta _{\varepsilon }$ is an isometry for the inner products on $%
\tilde{\Delta}(\varepsilon )$, i.e. 
\begin{equation}
\alpha _{i}\cdot \alpha _{j}=\tilde{\alpha}_{i}\cdot \tilde{\alpha}_{j},
\label{scalar}
\end{equation}%
which means 
\begin{equation}
\theta _{\varepsilon }^{\ast }=\theta _{\varepsilon }^{-1}\text{\qquad
and\qquad }\det \theta _{\varepsilon }=\pm 1.  \label{c3}
\end{equation}%
In summary, the task is to pick $\kappa $ elements of the Weyl group $\hat{%
\omega}_{i}$, expand the the deformation matrix $\theta _{\varepsilon }$ in
terms of the elements commuting with the products of these elements and
finally determined the coefficient functions $r_{k}(\varepsilon )$ in these
expansions for from the constraints 
\begin{equation}
\theta _{\varepsilon }^{\ast }\hat{\omega}_{i}=\hat{\omega}_{i}\theta
_{\varepsilon },\quad \left[ \hat{\omega}_{i}\hat{\omega}_{j},\theta
_{\varepsilon }\right] =0,\quad \theta _{\varepsilon }^{\ast }=\theta
_{\varepsilon }^{-1},\quad \det \theta _{\varepsilon }=\pm 1\quad \text{%
and\quad }\lim_{\varepsilon \rightarrow 0}\theta _{\varepsilon }=\mathbb{I}%
\text{,}  \label{const}
\end{equation}%
or possibly in reverse, that is for given $\theta _{\varepsilon }$ to
identify meaningful involutions $\hat{\omega}_{i}$. It turn out that these
constraints are quite restrictive and often allow to determine $\theta
_{\varepsilon }$ with only very few free parameters left. In some situations
it might not be desirable to preserve the inner products (\ref{scalar})
after the deformation, in which case one may give up (\ref{c3}).

With our applications to physical models of Calogero or Toda type in mind,
we may then easily construct a dual map $\delta ^{\star }$ for $\delta $ 
\begin{equation}
\delta ^{\star }:~\mathbb{R}^{n}\rightarrow \tilde{\Delta}^{\star
}(\varepsilon )=\mathbb{R}^{n}\oplus i \mathbb{R}^{n},\qquad x\mapsto \tilde{%
x}=\theta _{\varepsilon }^{\star }x,
\end{equation}%
i.e.~this map acts on the coordinate space with $x=\{x_{1},\ldots ,x_{n}\}$
or possibly fields. Given $\theta _{\varepsilon }$ we construct $\theta
_{\varepsilon }^{\star }$ by solving the $\ell $ equations 
\begin{equation}
(\tilde{\alpha}_{i}\cdot x)=(\left( \theta _{\varepsilon }\alpha \right)
_{i}\cdot x)=(\alpha _{i}\cdot \theta _{\varepsilon }^{\star }x)=(\alpha
_{i}\cdot \tilde{x}),\quad \text{for }i=1,\ldots ,\ell ,  \label{dual}
\end{equation}%
involving the standard inner product. This means $(\theta _{\varepsilon
}^{\star })^{-1}\alpha _{i}=\left( \theta _{\varepsilon }\alpha \right) _{i}$%
. Note that in general $\theta _{\varepsilon }^{\star }\neq \theta
_{\varepsilon }^{\ast }$. Naturally we can also identify an antilinear
involutory map 
\begin{equation}
\varpi ^{\star }:\tilde{\Delta}^{\star }(\varepsilon )\rightarrow \tilde{%
\Delta}^{\star }(\varepsilon ),\qquad \tilde{x}\mapsto \omega ^{\star }%
\tilde{x}.  \label{dom}
\end{equation}%
corresponding to $\varpi $ but acting in the dual space. Concretely we solve
for this the $\kappa \times \ell $ relations%
\begin{equation}
\left( \omega _{i}\tilde{\alpha}\right) _{j}\cdot x=\alpha _{j}\cdot \omega
_{i}^{\star }\tilde{x},\quad \text{for }i=1,\ldots \kappa \text{; }%
j=1,\ldots ,\ell ,  \label{xc}
\end{equation}%
for $\omega _{i}^{\star }$ with given $\omega _{i}$.

\section{Deformation matrices from factorised modified Coxeter elements}

As explained in the previous section, in principle the involution $\hat{%
\omega}_{i}$ could be \emph{any} element in the Weyl group. We will now
present a construction based on the selection of two specific, albeit still
fairly generic, elements $\hat{\omega}_{1}=\tilde{\sigma}_{-}$ and $\hat{%
\omega}_{2}=\tilde{\sigma}_{+}$ defined as 
\begin{equation}
\text{\ }\tilde{\sigma}_{\pm }:=\prod\limits_{i\in \tilde{V}_{\pm }}\sigma
_{i}.  \label{ti}
\end{equation}%
The $\sigma _{i}$ in (\ref{ti}) are simple Weyl reflections, acting as 
\begin{equation}
\sigma _{i}(x):=x-2\frac{x\cdot \alpha _{i}}{\alpha _{i}^{2}}\alpha
_{i},\qquad \text{with \ \ \ }1\leq i\leq \ell \equiv \limfunc{rank}\mathcal{%
W}.  \label{Weyl}
\end{equation}%
The sets $V_{\pm }$ are defined via the bi-colouration of the Dynkin diagram
as explained in \cite{FringSmith} and references therein. The difference
towards the treatment in \cite{FringSmith} is that the products in (\ref{ti}%
) do not have to extend over all possible elements in $V_{\pm }$, such that $%
\tilde{V}_{\pm }\subseteq V_{\pm }$. Denoting by $\sigma _{\pm }$ the
factors of $\tilde{\sigma}$ when $\tilde{h}=h$ we may therefore express the
reduced elements as $\tilde{\sigma}_{\pm }:=\sigma _{\pm
}\prod\nolimits_{j\in \breve{V}_{\pm }}\sigma _{i}$ for some values $j$,
which follows by recalling $[\sigma _{i},\sigma _{j}]=0$ for $i,j\in V_{+}$
or $i,j\in V_{-}$ and $\sigma _{i}^{2}=1$. Thus $\breve{V}_{\pm }$ is the
complement of \ $\tilde{V}_{\pm }$ in $V_{\pm }$, that is $V_{\pm }=$ $%
\breve{V}_{\pm }\cup \tilde{V}_{\pm }$. This ensures that we have maintained
the crucial involutory property $\tilde{\sigma}_{\pm }^{2}=1$.

This means the element $\Omega _{ij}$ in (\ref{ohm}) can be viewed as a
modified Coxeter element $\tilde{\sigma}:=\tilde{\sigma}_{-}\tilde{\sigma}%
_{+}$ with property 
\begin{equation}
\tilde{\sigma}^{\tilde{h}}=\mathbb{I},\qquad \ \ \ \ \ \ \ \text{with }%
\tilde{h}\leq h.
\end{equation}%
Therefore $\tilde{\sigma}$ equals a Coxeter element $\sigma $ when the order 
$\tilde{h}$ becomes the Coxeter number $h$.

The {reduced root space $\tilde{\Delta}$} is then constructed by acting with 
$\tilde{\sigma}$ on representatives {$\tilde{\gamma}_{i}=c_{i}\tilde{\alpha}%
_{i}$ of a particular orbit }$\tilde{\Omega}_{i}$ containing now $\tilde{h}${%
\ instead of }$h$ roots{\ 
\begin{equation}
\tilde{\Omega}_{i}:=\left\{ \gamma _{i},\tilde{\sigma}\gamma _{i},\tilde{%
\sigma}^{2}\gamma _{i},\ldots ,\tilde{\sigma}^{\tilde{h}-1}\gamma
_{i}\right\} .
\end{equation}%
The corresponding entire root space containing }$\ell \times \tilde{h}$
roots {is the union of all orbits%
\begin{equation}
\tilde{\Delta}=\bigcup\limits_{i=1}^{\ell }\tilde{\Omega}_{i}\text{.}
\end{equation}%
}

In analogy to the deformations defined in \cite{FringSmith} we construct
therefore the map $\varpi $ as 
\begin{equation}
\tilde{\sigma}_{\pm }^{\varepsilon }:=\theta _{\varepsilon }\tilde{\sigma}%
_{\pm }\theta _{\varepsilon }^{-1}=\tilde{\sigma}_{\pm }\tau  \label{c1}
\end{equation}%
where we assumed an additional property with $\theta _{\varepsilon }$ being
the deformation matrix as introduced in (\ref{defo}). Defining the deformed
reduced Coxeter element as $\tilde{\sigma}^{\varepsilon }:=\tilde{\sigma}%
_{-}^{\varepsilon }\tilde{\sigma}_{+}^{\varepsilon }$ we use a similar line
of reasoning as in the deduction of (\ref{ohm}) to show that $\left[ \tilde{%
\sigma},\theta _{\varepsilon }\right] =0$. Therefore we make the following
Ansatz for the deformation matrix%
\begin{equation}
\theta _{\varepsilon }=\sum\limits_{k=0}^{\tilde{h}-1}\mu _{k}(\varepsilon )%
\tilde{\sigma}^{k},\qquad \text{with }\lim_{\varepsilon \rightarrow 0}\mu
_{k}(\varepsilon )=\left\{ 
\begin{array}{l}
1\quad k=0 \\ 
0\quad k\neq 0%
\end{array}%
\right. ,~\mu _{k}(\varepsilon )\in \mathbb{C}.  \label{Ansatz}
\end{equation}%
The assumption for the coefficients $\mu _{k}(\varepsilon )$ ensures the
appropriate limit $\lim_{\varepsilon \rightarrow 0}\theta _{\varepsilon }=%
\mathbb{I}$. Equation (\ref{c1}) yields the constraint $\theta _{\varepsilon
}^{\ast }\tilde{\sigma}_{\pm }=\tilde{\sigma}_{\pm }\theta _{\varepsilon }$,
from which we deduce with (\ref{Ansatz}) 
\begin{equation}
\theta _{\varepsilon }=\left\{ 
\begin{array}{ll}
r_{0}(\varepsilon )\mathbb{I}+i\sum\limits_{k=1}^{(\tilde{h}%
-1)/2}r_{k}(\varepsilon )(\tilde{\sigma}^{k}-\tilde{\sigma}^{-k}) & \text{%
for }\tilde{h}\text{ odd,} \\ 
r_{0}(\varepsilon )\mathbb{I}+r_{\tilde{h}/2}(\varepsilon )\tilde{\sigma}^{%
\tilde{h}/2}+i\sum\limits_{k=1}^{\tilde{h}/2-1}r_{k}(\varepsilon )(\tilde{%
\sigma}^{k}-\tilde{\sigma}^{-k})~~~ & \text{for }\tilde{h}\text{ even,}%
\end{array}%
\right.  \label{th}
\end{equation}%
where $\mu _{0}(\varepsilon )=:r_{0}(\varepsilon )\in \mathbb{R},$ $\mu _{%
\tilde{h}/2}(\varepsilon )=:r_{\tilde{h}/2}(\varepsilon )\in \mathbb{R}$
when $\tilde{h}$ is even. In addition we defined $\mu _{k}(\varepsilon )=i
r_{k}(\varepsilon )$. Demanding next that $\theta _{\varepsilon }$ is an
isometry, we invoke the constraint $\det \theta _{\varepsilon }=1$. By means
of the eigenvalue equations for $\tilde{\sigma}$ 
\begin{equation}
\tilde{\sigma}\tilde{v}_{n}=e^{2\pi i\tilde{s}_{n}/\tilde{h}}\tilde{v}%
_{n}\qquad \ \ \text{with }n=1,\ldots \ell ,  \label{eig}
\end{equation}%
we define a set of \textquotedblleft modified exponents\textquotedblright\ $%
\tilde{s}=\{\tilde{s}_{1},\ldots ,\tilde{s}_{\ell }\}$. Unlike as for the
standard case, the eigenvalues may be degenerate in the modified scenario.
In general, they take on the values 
\begin{equation}
\tilde{s}=\left\{ 1^{\lambda _{1}},2^{\lambda _{2}},\ldots ,(\tilde{h}%
-1)^{\lambda _{\tilde{h}-1}},\tilde{h}^{\lambda _{\tilde{h}}}\right\} \qquad 
\text{with \ }\sum\limits_{k=1}^{\tilde{h}}\lambda _{k}=\ell ,  \label{mode}
\end{equation}%
with $\lambda _{i}$ indicating the degeneracy of certain eigenvalues in (\ref%
{eig}). Due to the degeneracy there could be several solutions to (\ref{eig}%
) with different elements $\tilde{\sigma}^{(i)}$ for $i=1,\ldots m$ forming
a similarity class 
\begin{equation}
\Sigma _{\tilde{s}}=\left\{ \tilde{\sigma}^{(1)}\tilde{\sigma}^{(2)},\ldots ,%
\tilde{\sigma}^{(m)}\right\} .
\end{equation}%
As in \cite{FringSmith} we demand the preservation of the inner product
between the original and deformed roots, which implies that $\det \theta
_{\varepsilon }=1$ and $\theta _{\varepsilon }^{\ast }=\theta _{\varepsilon
}^{-1}$. Diagonalising (\ref{th}) the constraint $\det \theta _{\varepsilon
}=1$ simply becomes 
\begin{equation}
\begin{array}{ll}
1=\prod\limits_{n=1}^{\ell }\left[ r_{0}(\varepsilon )-2\sum\limits_{k=1}^{(%
\tilde{h}-1)/2}r_{k}(\varepsilon )\sin \left( \frac{2\pi k}{\tilde{h}}\tilde{%
s}_{n}\right) \right] & \text{for }\tilde{h}\text{ odd,} \\ 
1=\prod\limits_{n=1}^{\ell }\left[ r_{0}(\varepsilon )+(-1)^{\tilde{s}%
_{n}}r_{\tilde{h}/2}(\varepsilon )-2\sum\limits_{k=1}^{\tilde{h}%
/2-1}r_{k}(\varepsilon )\sin \left( \frac{2\pi k}{\tilde{h}}\tilde{s}%
_{n}\right) \right] ~~~ & \text{for }\tilde{h}\text{ even.}%
\end{array}
\label{e1}
\end{equation}%
Solving these constraints for $\theta _{\varepsilon }$ allows us to
construct the simple roots $\tilde{\alpha}_{i}$ and therefore the entire
deformed {reduced root space $\tilde{\Delta}(\varepsilon )$. Note that for
simplicity we use the same notation for the unformed and deformed root
space, distinguishing the latter always by the explicit mentioning of the
deformation parameter }$\varepsilon $.{\ Hence we have 
\begin{equation}
\tilde{\Omega}_{i}^{\varepsilon }=\theta _{\varepsilon }\tilde{\Omega}_{i},
\label{omega}
\end{equation}%
and therefore%
\begin{equation}
\tilde{\Delta}(\varepsilon )=\bigcup\limits_{i=1}^{\ell }\tilde{\Omega}%
_{i}^{\varepsilon }=\theta _{\varepsilon }\tilde{\Delta}\text{.}
\label{omega2}
\end{equation}%
This construction guarantees that the }$\tilde{\sigma}_{\pm }^{\varepsilon }$
are indeed representations of the map $\varpi $ in (\ref{map}). Evidently it
leaves the root space invariant {\ 
\begin{equation}
\tilde{\sigma}_{\pm }^{\varepsilon }:\tilde{\Delta}(\varepsilon )\rightarrow
\theta _{\varepsilon }\tilde{\sigma}_{\pm }\theta _{\varepsilon }^{-1}\tilde{%
\Delta}(\varepsilon )=\theta _{\varepsilon }\tilde{\sigma}_{\pm }\tilde{%
\Delta}=\theta _{\varepsilon }\tilde{\Delta}=\tilde{\Delta}(\varepsilon ).
\label{a}
\end{equation}%
For the latter property to hold we may also exclude some of the orbits }$%
\tilde{\Omega}_{i}^{\varepsilon }$ {in the union} $\bigcup\nolimits_{i=1}^{%
\ell },$ whenever they are mapped into themselves $\tilde{\sigma}_{\pm
}^{\varepsilon }:\tilde{\Omega}_{i}^{\varepsilon }\rightarrow \tilde{\Omega}%
_{i}^{\varepsilon }$.

\subsection{Antilinearly deformed $A_{\ell }$ root systems}

When engaging into a case-by-case description in \cite{FringSmith}, we
characterized different solutions group by group. Here we will take equation
(\ref{e1}) as more fundamental and classify the solutions according to
different values of the modified Coxeter number. In this manner different
types of solutions to (\ref{e1}) are then characterized by different sets of
modified exponents (\ref{mode}). This means we need to verify subsequently
whether a corresponding $\tilde{\sigma}$ really exists.

For definiteness we fix our conventions and associate the colour values $%
c_{i}=1$ or $c_{i}=-1$ when $i$ is even or odd, respectively, to the
vertices of the Dynkin diagram. We find various similarity classes $\Sigma _{%
\tilde{s}}$ characterized by different sets of modified exponents $\tilde{s}$%
.

\subsubsection{The class with modified exponents $\{{1,2,3,4^{\ell -3}\}}$
and \~{h}=4}

We find that the simplest similarity class $\Sigma $ for which $x^{4}=1$
when $x\in \Sigma $ is 
\begin{equation}
\Sigma _{\{1,2,3,4^{\ell -3}\}}=\left\{ \tilde{\sigma}^{(1)},\ldots ,\tilde{%
\sigma}^{(\ell -2)}\right\} ,  \label{cl1}
\end{equation}%
where the elements of that class are defined as%
\begin{equation}
\tilde{\sigma}^{(i)}:=\left( \sigma _{i+1}\sigma _{i}\sigma _{i+2}\right)
^{c_{i}}\text{\qquad for }i=1,\ldots ,\ell -2.  \label{si}
\end{equation}

It is clear that each element $\tilde{\sigma}^{(i)}$ in (\ref{si}) has order 
$4$, since it is formed from three consecutive elements on the Dynkin
diagram and thus being isomorphic to the Coxeter element of {$A_{3}$ when
acting on the three corresponding roots.}

Furthermore, by definition all elements of $\Sigma $ have to be related by a
similarity transformation. Indeed we find: \emph{Two consecutive elements in 
}$\Sigma _{\{1,2,3,4^{\ell -3}\}}$ \emph{are related as}%
\begin{equation}
\varkappa _{i}\tilde{\sigma}^{(i)}=\tilde{\sigma}^{(i+1)}\varkappa _{i}\text{%
\qquad with }\varkappa _{i}:=\sigma _{i}\sigma _{i+1}\sigma _{i+2}\sigma
_{i+3}\sigma _{i+1}.  \label{sim}
\end{equation}%
\emph{Therefore all elements in }$\Sigma $\emph{\ can be related to each
other by an adjoint action simply by successive applications of (\ref{sim}).}

Proof: Let us now prove the relation (\ref{sim}). The starting point is the
identity%
\begin{equation}
\sigma _{i-1}\sigma _{i}\sigma _{i+1}\sigma _{i}=\sigma _{i+1}\sigma
_{i-1}\sigma _{i}\sigma _{i+1},  \label{ident}
\end{equation}%
which follows by applying the left and right hand side to some arbitrary $x$
using the definition of the simple Weyl reflection (\ref{Weyl})
consecutively. Normalising the length of the roots to be $2$, we find in
both cases%
\begin{equation}
x-\left[ (x\cdot \alpha _{i-1})+(x\cdot \alpha _{i})+(x\cdot \alpha _{i+1})%
\right] \alpha _{i-1}-\left[ (x\cdot \alpha _{i})+(x\cdot \alpha _{i+1})%
\right] (\alpha _{i}+\alpha _{i+1}).
\end{equation}%
Multiplying (\ref{ident}) from the left by $\prod\nolimits_{k=1}^{i-2}\sigma
_{k}$ and $\prod\nolimits_{k=i+2}^{\ell }\sigma _{k}$ from the right and
noting that for\ $A_{\ell }$ we have $[\sigma _{i},\sigma _{j}]=0$ for $%
\left\vert i-j\right\vert \geq 2$, it follows%
\begin{equation}
\hat{\sigma}\sigma _{i}=\sigma _{i+1}\hat{\sigma},\text{\qquad with }\hat{%
\sigma}:=\prod\limits_{k=1}^{\ell }\sigma _{k}.  \label{ss}
\end{equation}%
The element $\hat{\sigma}$ is the standard Coxeter element. Multiplying next
the identity (\ref{sim}) from the left by $\prod\nolimits_{k=1}^{i-1}\sigma
_{k}$ and $\sigma _{i+1}\prod\nolimits_{k=i+4}^{\ell }\sigma _{k}$ from the
right and recalling that $\sigma _{i}^{2}=1$ yields%
\begin{equation}
\hat{\sigma}\left( \sigma _{i}\sigma _{i+2}\sigma _{i+1}\right)
^{c_{i}}=\left( \sigma _{i+1}\sigma _{i+3}\sigma _{i+2}\right) ^{c_{i}}\hat{%
\sigma}.
\end{equation}%
This relation is now easily established by commuting all three simple Weyl
reflections through the Coxeter element using the identity (\ref{ss}), which
in turn also proves (\ref{sim}). $\square $

Some special elements in $\Sigma $ are related by the adjoint action of the
Coxeter element $\sigma $. We find: \emph{The first and the last element in }%
$\Sigma _{\{1,2,3,4^{\ell -3}\}}$\emph{\ are related as}%
\begin{equation}
\text{ }\tilde{\sigma}^{(\ell -2)}\sigma ^{\frac{h-c_{\ell }}{2}}=\sigma ^{%
\frac{h-c_{\ell }}{2}}\tilde{\sigma}^{(1)},  \label{rel}
\end{equation}%
Proof: We prove (\ref{rel}) by using the more elementary relations%
\begin{equation}
\sigma _{\ell +1-i}\sigma ^{\frac{h}{2}+\frac{c_{i}+c_{i}c_{\ell }}{4}%
}=\sigma ^{\frac{h}{2}+\frac{c_{i}+c_{i}c_{\ell }}{4}}\sigma _{i}.
\label{ew}
\end{equation}%
For even $h$ we compute by a successive use of (\ref{ew})%
\begin{equation}
\text{ }\tilde{\sigma}^{(\ell -2)}\sigma ^{\frac{h}{2}}=\sigma _{\ell
-2}\sigma _{\ell }\sigma _{\ell -1}\sigma ^{\frac{h}{2}}=\sigma _{\ell
-2}\sigma _{\ell }\sigma ^{\frac{h}{2}}\sigma _{2}=\sigma _{\ell -2}\sigma ^{%
\frac{h}{2}}\sigma _{1}\sigma _{2}=\sigma ^{\frac{h}{2}}\sigma _{3}\sigma
_{1}\sigma _{2}=\sigma ^{\frac{h}{2}}\tilde{\sigma}^{(1)}.
\end{equation}%
Similarly we compute for odd $h$ 
\begin{eqnarray}
\text{ }\tilde{\sigma}^{(\ell -2)}\sigma ^{\frac{h-1}{2}} &=&\sigma _{\ell
-1}\sigma _{\ell -2}\sigma _{\ell }\sigma ^{\frac{h-1}{2}}=\sigma _{\ell
-1}\sigma _{\ell -2}\sigma ^{\frac{h-1}{2}}\sigma _{1}=\sigma _{\ell
-1}\sigma ^{\frac{h-1}{2}}\sigma _{3}\sigma _{1} \\
&=&\sigma _{\ell -1}\sigma ^{\frac{h+1}{2}}\sigma ^{-1}\sigma _{3}\sigma
_{1}=\sigma ^{\frac{h+1}{2}}\sigma _{2}\sigma ^{-1}\sigma _{3}\sigma
_{1}=\sigma ^{\frac{h-1}{2}}\sigma _{-}\sigma _{+}\sigma _{2}\sigma
_{+}\sigma _{-}\sigma _{3}\sigma _{1}  \notag \\
&=&\sigma ^{\frac{h-1}{2}}\sigma _{3}\sigma _{1}\sigma _{2}=\sigma ^{\frac{%
h-1}{2}}\tilde{\sigma}^{(1)}.  \notag
\end{eqnarray}%
Thus we have established that the first element $\tilde{\sigma}^{(1)}$ in
the similarity class $\Sigma $ is related via the similarity transformation (%
\ref{rel}) to the last element $\tilde{\sigma}^{(\ell -2)}$ in this class.
In comparison to one rank less the last element is the only additional one.
For the other elements we can use the same argumentation but employing the
Coxeter element for one rank less. $\square $

\paragraph{Expl.: {$A_{8}$}}

We illustrate now the working of these formulae for a concrete{\ example. We
consider $A_{8}$ and generate the entire root space $\tilde{\Delta}$ as
described in (\ref{omega}) from $\tilde{\sigma}^{(1)}$. The results are
depicted in table 1. }

\begin{table}[h!]
\begin{center}
\begin{tabular}{|c|c|c|c|c|c|c|c|c|}
\hline
$( \tilde{\sigma}^{(1)} )^{j}\backslash \alpha _{i}$ & $\alpha _{1}$ & $%
\alpha _{2}$ & $\alpha _{3}$ & $\alpha _{4}$ & $\alpha _{5}$ & $\alpha _{6}$
& $\alpha _{7}$ & $\alpha _{8}$ \\ \hline
$\tilde{\sigma}^{(1)}$ & $\mathbf{-1,2}$ & $\mathbf{1,2,3}$ & $\mathbf{-2,3}$
& $2,3,4$ & $5$ & $6$ & $7$ & $8$ \\ \hline
$\tilde{\sigma}^{(1)}\tilde{\sigma}^{(1)}$ & $\mathbf{-3}$ & $\mathbf{-2}$ & 
$\mathbf{-1}$ & $1,2,3,4$ & $5$ & $6$ & $7$ & $8$ \\ \hline
$\tilde{\sigma}^{(1)}\tilde{\sigma}^{(1)}\tilde{\sigma}^{(1)}$ & $\mathbf{2,3%
}$ & $\mathbf{-1,2,3}$ & $\mathbf{1,2}$ & $3,4$ & $5$ & $6$ & $7$ & $8$ \\ 
\hline
\end{tabular}%
\end{center}
\caption{The reduced $A_{8}$-root space $\tilde{\Delta}$ generated from the
orbits of $\tilde{\protect\sigma}^{(1)}$. }
\end{table}

For convenience we used the following conventions: For any non-simple root $%
\beta =\sum\nolimits_{i}\mu _{i}\alpha _{i}$ we present only the
non-vanishing coefficients $\mu _{i}$ in the table with the overall sign
written in front, e.g.~$\alpha _{1}+\alpha _{2}+\alpha _{3}$ is represented
as $1,2,3$ and $-\alpha _{1}-\alpha _{2}$ as$~-1,2$. We indicate the $A_{3}$
substructure in bold. Further examples for root spaces obtained from
different elements in $\Sigma _{\{1,2,3,4^{\ell -3}\}}$ are presented in
appendix A.

Crucial to our construction is the invariance under the action of $\tilde{%
\sigma}_{\pm }^{(1)}$. Acting on the roots as depicted in table 1 with $%
\tilde{\sigma}_{\pm }^{(1)}$ we recover all the elements in table 1, albeit
in a permuted way as indicated in table 2.

\begin{table}[h!]
\begin{center}
\begin{tabular}{|c|c|c|c|c|c|c|c|c|}
\hline
$\tilde{\sigma}_{-}^{(1)}( {\tilde{\Delta}}) $ & $-1$ & $1,2,3$ & $-3$ & $%
3,4 $ & $5$ & $6$ & $7$ & $8$ \\ \hline
& $-2,3$ & $2$ & $-1,2$ & $1,2,3,4$ & $5$ & $6$ & $7$ & $8$ \\ \hline
& $3$ & $-1,2,3$ & $-1$ & $2,3,4$ & $5$ & $6$ & $7$ & $8$ \\ \hline
& $1,2$ & $-2$ & $2,3$ & $4$ & $5$ & $6$ & $7$ & $8$ \\ \hline
\end{tabular}%
\par
\medskip 
\begin{tabular}{|c|c|c|c|c|c|c|c|c|}
\hline
$\tilde{\sigma}_{+}^{(1)}( {\tilde{\Delta}}) $ & $1,2$ & $-2$ & $2,3$ & $4$
& $5$ & $6$ & $7$ & $8$ \\ \hline
& $-1$ & $1,2,3$ & $-3$ & $3,4$ & $5$ & $6$ & $7$ & $8$ \\ \hline
& $-2,3$ & $2$ & $-1,2$ & $1,2,3,4$ & $5$ & $6$ & $7$ & $8$ \\ \hline
& $3$ & $-1,2,3$ & $1$ & $3,4$ & $5$ & $6$ & $7$ & $8$ \\ \hline
\end{tabular}%
\end{center}
\caption{The invariance of the $A_{8}$-root space $\tilde{\Delta}$ generated
from $\tilde{\protect\sigma}^{(1)}$ under the action of $\tilde{\protect%
\sigma}_{\pm}^{(1)}$.}
\end{table}

\subsubsection{The class with modified exponents $\left\{ {1,2^{2},3,4^{\ell
-3}}\right\} $ and \~{h}=4}

Other classes become considerably more complicated. We present here only
some examples to indicate this. For instance in the class 
\begin{equation}
\Sigma _{\{1,2^{2},3,4^{\ell -4}\}}=\left\{ \tilde{\sigma}^{(1,1,1)},\tilde{%
\sigma}^{(1,1,2)},\ldots ,\tilde{\sigma}^{(2,1,\ell -4)}\right\} ,
\label{class2}
\end{equation}%
we have to label the elements by three indices%
\begin{equation}
\tilde{\sigma}^{(1,i,j)}:=\sigma _{i}\sigma _{i+2}\sigma _{i+3+j}\sigma
_{i+1}\qquad \text{and\qquad }\tilde{\sigma}^{(2,i,j)}:=\sigma _{i}\sigma
_{i+1+j}\sigma _{i+3+j}\sigma _{i+j+2}  \label{sti21}
\end{equation}%
with $i=1,\ldots ,\ell -j-3$ and $j=1,\ldots ,\ell -4$. It is easy to
convince oneself that these elements have order $4$. In both types of
labeling we have three consecutive elements and one additional factor which
commutes with all the other elements, that is $\sigma _{i+3+j}$ in $\tilde{%
\sigma}^{(1,i,j)}$ and $\sigma _{i}$ in $\tilde{\sigma}^{(2,i,j)}$,
respectively. Thus by the same argument as in the previous class and the
fact that $\sigma _{i}^{2}=1$ it follows that the order of all elements in (%
\ref{sti21}) is $4$.

Arguing along similar lines as for the class presented in the previous
subsection, we can also show that all elements in $\Sigma
_{\{1,2^{2},3,4^{\ell -4}\}}$ are indeed related by a similarity
transformation. We will not present this proof here.

\subsubsection{The similarity class structure with \~{h}=4}

It is clear that for higher ranks more and more possible sets of exponents
characterising different classes may exist. Here we only indicate in table 3
the general structure but do not report a detailed construction of the
elements of these classes and their interrelations as the argumentation goes
along the same lines as in the two previous subsections. By inspection of
the table we notice the onset of two new classes when we increase the rank
by two, that is the number of classes increases by $2$ for $\ell =2n+5$ for $%
n=1,2,\ldots $ We also observe that the number of classes for $\ell =2n+1$
and $\ell =2n+2$ is the same.

{\renewcommand{\tabcolsep}{0.05cm} 
\begin{table}[h!]
\begin{tabular}{|c|c|c|c|c|c|c|}
\hline
$\ell$ &  &  &  &  &  &  \\ \hline
$3$ & $\{1,2,3 \}$ &  &  &  &  &  \\ \hline
$4$ & $\{1,2,3,4\}$ &  &  &  &  &  \\ \hline
$5$ & $\{1,2,3,4^2\}$ & $\{1,2^2,3,4\}$ &  &  &  &  \\ \hline
$6$ & $\{1,2,3,4^3\}$ & $\{1,2^2,3,4^2\}$ &  &  &  &  \\ \hline
$7$ & $\{1,2,3,4^4\}$ & $\{1,2^2,3,4^3\}$ & $\{1,2^3,3,4^2\}$ & $%
\{1^2,2^2,3^2,4\}$ &  &  \\ \hline
$8$ & $\{1,2,3,4^5\}$ & $\{1,2^2,3,4^4\}$ & $\{1,2^3,3,4^3\}$ & $%
\{1^2,2^2,3^2,4^2\}$ &  &  \\ \hline
$9$ & $\{1,2,3,4^6\}$ & $\{1,2^2,3,4^5\}$ & $\{1,2^3,3,4^4\}$ & $%
\{1^2,2^2,3^2,4^3\}$ & $\{1^2,2^3,3^2,4^2\}$ & $\{1,2^4,3,4^3\}$ \\ \hline
$10$ & $\{1,2,3,4^7\}$ & $\{1,2^2,3,4^6\}$ & $\{1,2^3,3,4^5\}$ & $%
\{1^2,2^2,3^2,4^4\}$ & $\{1^2,2^3,3^2,4^3\}$ & $\{1,2^4,3,4^4\}$ \\ \hline
\vdots & \vdots & \vdots & \vdots & \vdots & \vdots & \ldots \\ \hline
$\ell$ & $\{1,2,3,4^{\ell-3}\}$ & $\{1,2^2,3,4^{\ell-4}\}$ & $%
\{1,2^3,3,4^{\ell-5}\}$ & $\{1^2,2^2,3^2,4^{\ell-6}\}$ & $%
\{1^2,2^3,3^2,4^{\ell-7}\}$ & \ldots \\ \hline
\end{tabular}%
\caption{Similarity classes in $A_{\ell}$ with $\tilde{h}=4$.}
\end{table}
}

In addition we note that the number of factors in the elements of a
similarity class increases by one in the table in each column from the left
to the right, starting with three factors on the very left.

\subsubsection{The class with modified exponents $\left\{ 1,2,\ldots
,4n-1,4n^{\ell -4n+1}\right\} $ and \~{h}=4n}

Let us now generalize the previous considerations towards classes with
larger amounts of eigenvalues, such that they are related to modified
Coxeter numbers of higher powers. The class (\ref{cl1}) acquires the more
general form 
\begin{equation}
\Sigma _{\{1,2,\ldots ,4n-1,4n^{\ell -4n+1}\}}=\left\{ \tilde{\sigma}%
^{(n,1)},\ldots ,\tilde{\sigma}^{(n,\ell +2-4n)}\right\} ,
\end{equation}%
when $x^{4n}=1$ for $x\in \Sigma $. In this case the elements of the class 
\begin{equation}
\tilde{\sigma}^{(n,i)}:=\left[ \left( \prod\limits_{k=2}^{n}\sigma
_{i-1+4(k-1)}\sigma _{i+1+4(k-1)}\right) \sigma _{i+1}\left(
\prod\limits_{k=1}^{n}\sigma _{i+4(k-1)}\sigma _{i+2+4(k-1)}\right) \right]
^{c_{i}},  \label{cl1x}
\end{equation}%
are characterised by two indices, $n$ distinguishing the particular type of
class and $i=1,\ldots ,\ell +2-4n$ labeling the individual elements in that
class. The case $n=1$ reduces to our previous simpler example with $\tilde{%
\sigma}^{(n,i)}=\tilde{\sigma}^{(i)}$ as defined in (\ref{si}). Evidently
the element $\tilde{\sigma}^{(n,i)}$ contains the $4n-1$ consecutive factors 
$\sigma _{i}$ to $\sigma _{i+4n-3}$ separated into odd and even indices.
This means each element can be viewed as a Coxeter element for the {$%
A_{4n-1} $-Weyl group and therefore the order of }$\tilde{\sigma}^{(n,i)}$
is\ {$\tilde{h}=4n$. }

In this case we will also establish that all elements in $\Sigma $ are
indeed related by a similarity transformation. Two consecutive elements in
this class are related as%
\begin{equation}
\varkappa _{i}^{(n)}\tilde{\sigma}^{(n,i)}=\tilde{\sigma}^{(n,i+1)}\varkappa
_{i}^{(n)}\text{\qquad with }\varkappa
_{i}^{(n)}:=\prod\limits_{k=1}^{4n}\sigma
_{i+k-1}\prod\limits_{k=1}^{2n-1}\sigma _{i+2k-1},
\end{equation}%
which in turn implies that all elements in $\Sigma $ are related by a
similarity transformation. The proof for this identity goes along the same
line as the one for the particular case $n=1$ of the identity (\ref{sim}).

\subsubsection{The class with modified exponents $\left\{ 1,2^{2},\ldots
,4n-1,4n^{\ell -4n}\right\} $ and \~{h}=4n}

For higher order the similarity class (\ref{class2}) generalises to%
\begin{equation}
\Sigma _{\{1,2^{2},\ldots ,4n-1,4n^{\ell -4n+1}\}}=\left\{ \tilde{\sigma}%
^{(1,1,1,1)},\tilde{\sigma}^{(1,2,1,1)},\ldots \right\} ,
\end{equation}%
where we label its elements 
\begin{eqnarray}
\tilde{\sigma}^{(1,n,i,j)} &:&=\prod\limits_{k=1}^{n}\sigma
_{i+4(k-1)}\sigma _{i+2+4(k-1)}\sigma _{i+j+(\tilde{h}-1)}\sigma
_{i+1}\prod\limits_{k=2}^{n}\sigma _{i-1+4(k-1)}\sigma _{i+1+4(k-1),}\quad
\label{e34} \\
\tilde{\sigma}^{(2,n,i,j)} &:&=\sigma _{i+j+2}\prod\limits_{k=1}^{n}\sigma
_{i+j+4(k-1)}\sigma _{i+j+2+4(k-1)}\sigma _{i}\prod\limits_{k=2}^{n}\sigma
_{i+j+1+4(k-1)}\sigma _{i+j+3+4(k-1)},~~~~  \label{e35}
\end{eqnarray}%
now by four indices with $j=1,\cdots ,\ell -4n$ and $i=1,\cdots ,\ell
-j-(4n-1)$. We recover the case discussed in the previous section for $n=1$.
Using similar arguments as before we can show that all elements in this
class have order \~{h}=4n. For instance for the element $\sigma _{i+j+(%
\tilde{h}-1)}$ in (\ref{e34}) the subscript obeys $i+j+(\tilde{h}-1)>i+%
\tilde{h}$, which means that the element may be commuted to the left. Taking
then the $\tilde{h}$-th power of the entire expression we find

\begin{equation}
\Big[(\sigma _{i+j+(\tilde{h}-1)})^{\tilde{h}}\big[\Big(\prod%
\limits_{k=1}^{n}\sigma _{i+4(k-1)}\sigma _{i+2+4(k-1)}\Big)\sigma _{i+1}%
\Big(\prod\limits_{k=2}^{n}\sigma _{i-1+4(k-1)}\sigma _{i+1+4(k-1)}\Big)\big]%
^{\tilde{h}}\Big].  \label{36}
\end{equation}%
Since $\tilde{h}$ is even we have $(\sigma _{i+j+(\tilde{h}-1)})^{\tilde{h}%
}=1$ and since the expression in the bracket is a reduced Coxeter element
for $A_{\tilde{h}=4n}$ the expression in (\ref{36}) equals 1, thus
establishing the order of $\tilde{\sigma}^{(1,n,i,j)}$ to be $\tilde{h}=4n\,$%
. Similar arguments can be used for $\tilde{\sigma}^{(2,n,i,j)}$ to prove
that this element has the same order.

\subsubsection{Antilinearly invariant complex root spaces}

Based on the various classes constructed in the previous sections we may now
compute the deformation matrix with the help of the Ansatz (\ref{th})
subject to the mentioned constraints. In \cite{FringSmith} we found some
relatively simple solutions for $h=4n$. We present now similar solutions for 
$\tilde{h}=4n$. Taking in (\ref{e1}) all but three coefficients to be zero%
\begin{equation}
r_{i}(\varepsilon )=0\qquad \text{for }i\neq 0,n,2n,
\end{equation}%
the equation reduces with the help of (\ref{mode}) to 
\begin{equation}
1=(r_{0}+r_{2n})^{2\sum\nolimits_{k=1}^{n}\lambda _{2k}}\left[
(r_{0}-r_{2n})^{2}-4r_{n}^{2}\right] ^{\sum\nolimits_{k=1}^{n}\lambda
_{2k-1}}.
\end{equation}%
As can be seen directly, this equation is solved by%
\begin{equation}
r_{2n}=1-r_{0}\qquad \text{and\qquad }r_{n}=\sqrt{r_{0}(r_{0}-1)}=:\vartheta
.  \label{rr}
\end{equation}%
Thus the corresponding deformation matrix resulting from (\ref{th}) reads%
\begin{equation}
\theta _{\varepsilon }=r_{0}(\varepsilon )\mathbb{I}+\left[
1-r_{0}(\varepsilon )\right] \tilde{\sigma}^{2n}+i\vartheta (\tilde{\sigma}%
^{n}-\tilde{\sigma}^{-n}).  \label{xcv}
\end{equation}%
All what is left now is to establish whether the set of modified exponents
in (\ref{mode}) really exists for some concrete elements of $\tilde{\sigma}%
\in \mathcal{W}$ of order $\tilde{h}=4n$ and possibley to specify the
function $r_{0}(\varepsilon )$.

It is useful to consider a concrete example. For instance, the deformed
roots resulting from $\tilde{\sigma}^{(3)}$ of the class $\Sigma
_{\{1,2,3,4^{\ell -3}\}}$ for $A_{8}$ according to (\ref{xcv}) are%
\begin{equation}
\begin{array}{l}
\tilde{\alpha}_{1}=\alpha _{1},~\tilde{\alpha}_{7}=\alpha _{7},~\tilde{\alpha%
}_{8}=\alpha _{8}, \\ 
\tilde{\alpha}_{2}=\alpha _{2}+(1-\text{$r_{0}$})\alpha _{3}+(1-\text{$r_{0}$%
}+i\vartheta )\alpha _{4}+(1-\text{$r_{0}$})\alpha _{5}, \\ 
\tilde{\alpha}_{3}=(\text{$r_{0}$}-i\vartheta )\alpha _{3}-2i\vartheta
\alpha _{4}+(\text{$r_{0}$}-i\vartheta -1)\alpha _{5}, \\ 
\tilde{\alpha}_{4}=2i\vartheta \alpha _{3}+(2\text{$r_{0}$}+2i\vartheta
-1)\alpha _{4}+2i\vartheta \alpha _{5}, \\ 
\tilde{\alpha}_{5}=(\text{$r_{0}$}-i\vartheta -1)\alpha _{3}-2i\vartheta
\alpha _{4}+(\text{$r_{0}$}-i\vartheta )\alpha _{5}, \\ 
\tilde{\alpha}_{6}=(1-\text{$r_{0}$})\alpha _{3}+(1-\text{$r_{0}$}%
+i\vartheta )\alpha _{4}+(1-\text{$r_{0}$})\alpha _{5}+\alpha _{6}.%
\end{array}
\label{234}
\end{equation}%
The $\theta _{\varepsilon }$ resulting from different elements in the same
class have a similar form with the $A_{3}$-substructure displaced similarly
as for the undeformed roots. We do not report these solutions here. Unlike
as in (\ref{234}) all eight roots are deformed when constructing $\theta
_{\varepsilon }$ for instance from $\tilde{\sigma}^{(2,1)}$ as specified in (%
\ref{cl1x})%
\begin{equation}
\theta _{\varepsilon }=\left( 
\begin{array}{cccccccc}
\text{$r_{0}$} & 0 & i\vartheta & 2i\vartheta & i\vartheta & 0 & \text{$%
r_{0} $}-1 & 0 \\ 
0 & \text{$r_{0}$}-i\vartheta & -2i\vartheta & -2i\vartheta & -2i\vartheta & 
\text{$r_{0}$}-i\vartheta -1 & 0 & 0 \\ 
i\vartheta & 2i\vartheta & \text{$r_{0}$}+2i\vartheta & 2i\vartheta & \text{$%
r_{0}$}+2i\vartheta -1 & 2i\vartheta & i\vartheta & 0 \\ 
-2i\vartheta & -2i\vartheta & -2i\vartheta & 2\text{$r_{0}$}-2i\vartheta -1
& -2i\vartheta & -2i\vartheta & -2i\vartheta & 0 \\ 
i\vartheta & 2i\vartheta & \text{$r_{0}$}+2i\vartheta -1 & 2i\vartheta & 
\text{$r_{0}$}+2i\vartheta & 2i\vartheta & i\vartheta & 0 \\ 
0 & \text{$r_{0}$}-i\vartheta -1 & -2i\vartheta & -2i\vartheta & -2i\vartheta
& \text{$r_{0}$}-i\vartheta & 0 & 0 \\ 
\text{$r_{0}$}-1 & 0 & i\vartheta & 2i\vartheta & i\vartheta & 0 & \text{$%
r_{0}$} & 0 \\ 
1-\text{$r_{0}$} & 1-\text{$r_{0}$} & 1-\text{$r_{0}$} & 1-\text{$r_{0}$}%
-i\vartheta & 1-\text{$r_{0}$} & 1-\text{$r_{0}$} & 1-\text{$r_{0}$} & 1%
\end{array}%
\right) .  \label{theta}
\end{equation}%
The dual map $\delta ^{\star }$ is obtained by solving (\ref{dual}) for the
dual deformation matrix $\theta _{\varepsilon }^{\star }$ with the explicit
form for $\theta _{\varepsilon }$. Taking the latter to be given by (\ref%
{theta}) we compute for the standard $(\ell +1)$-dimensional representation
of $A_{\ell }$ $(\alpha _{i})_{j}=\delta _{ij}-\delta _{(i+1)j}$, $%
i=1,2,\ldots ,\ell $, $j=1,2,\ldots ,\ell +1$%
\begin{equation}
\theta _{\varepsilon }^{\star }=\left( 
\begin{array}{ccccccccc}
\text{$r_{0}$} & 0 & 0 & i\vartheta & -i\vartheta & 0 & 0 & 1-\text{$r_{0}$}
& 0 \\ 
0 & \text{$r_{0}$} & -i\vartheta & 0 & 0 & i\vartheta & 1-\text{$r_{0}$} & 0
& 0 \\ 
0 & i\vartheta & \text{$r_{0}$} & 0 & 0 & 1-\text{$r_{0}$} & -i\vartheta & 0
& 0 \\ 
-i\vartheta & 0 & 0 & \text{$r_{0}$} & 1-\text{$r_{0}$} & 0 & 0 & i\vartheta
& 0 \\ 
i\vartheta & 0 & 0 & 1-\text{$r_{0}$} & \text{$r_{0}$} & 0 & 0 & -i\vartheta
& 0 \\ 
0 & -i\vartheta & 1-\text{$r_{0}$} & 0 & 0 & \text{$r_{0}$} & i\vartheta & 0
& 0 \\ 
0 & 1-\text{$r_{0}$} & i\vartheta & 0 & 0 & -i\vartheta & \text{$r_{0}$} & 0
& 0 \\ 
1-\text{$r_{0}$} & 0 & 0 & -i\vartheta & i\vartheta & 0 & 0 & \text{$r_{0}$}
& 0 \\ 
0 & 0 & 0 & 0 & 0 & 0 & 0 & 0 & 1%
\end{array}%
\right) .
\end{equation}%
By construction the corresponding dual root space $\tilde{\Delta}^{\star
}(\varepsilon )$ is invariant under the action of some antilinear maps $%
\varpi ^{\star }$, obtained by solving (\ref{xc}). For antilinear symmetry $%
\omega _{1}=\tau \sigma _{2}\sigma _{4}\sigma _{6}$ we compute the dual
antilinear transformation to%
\begin{equation}
\omega _{1}^{\star }=\tau \left( 
\begin{array}{ccccccccc}
\nu ^{2} & 0 & 0 & -2i\vartheta \nu & 2i\vartheta \nu & 0 & 0 & \mu & 0 \\ 
0 & -2i\vartheta \nu & \nu ^{2} & 0 & 0 & \mu & 2i\vartheta \nu & 0 & 0 \\ 
0 & \nu ^{2} & 2i\vartheta \nu & 0 & 0 & -2i\vartheta \nu & \mu & 0 & 0 \\ 
-2i\vartheta \nu & 0 & 0 & \mu & \nu ^{2} & 0 & 0 & 2i\vartheta \nu & 0 \\ 
2i\vartheta \nu & 0 & 0 & \nu ^{2} & \mu & 0 & 0 & -2i\vartheta \nu & 0 \\ 
0 & \mu & -2i\vartheta \nu & 0 & 0 & 2i\vartheta \nu & \nu ^{2} & 0 & 0 \\ 
0 & 2i\vartheta \nu & \mu & 0 & 0 & \nu ^{2} & -2i\vartheta \nu & 0 & 0 \\ 
\mu & 0 & 0 & 2i\vartheta \nu & -2i\vartheta \nu & 0 & 0 & \nu ^{2} & 0 \\ 
0 & 0 & 0 & 0 & 0 & 0 & 0 & 0 & 1%
\end{array}%
\right) ,  \label{star1}
\end{equation}%
where we abbreviated $\nu :=2r_{0}$$-1$ and $\mu :=4(r_{0}$$-r_{0}^{2}$$)$.
Then the action on the deformed and original variables amounts with (\ref%
{star1}) simply to 
\begin{eqnarray}
\omega _{1}^{\star }:\tilde{\Delta}^{\ast }(\varepsilon )\rightarrow \tilde{%
\Delta}^{\ast }(\varepsilon ), &&\tilde{x}_{1}\mapsto \tilde{x}_{1},\tilde{x}%
_{2}\leftrightarrow \tilde{x}_{3},\tilde{x}_{4}\leftrightarrow \tilde{x}%
_{5},,\tilde{x}_{6}\leftrightarrow \tilde{x}_{7},\tilde{x}_{8}\mapsto \tilde{%
x}_{8},\tilde{x}_{9}\mapsto \tilde{x}_{9}, \\
&&x_{1}\mapsto x_{1},x_{2}\leftrightarrow x_{3},x_{4}\leftrightarrow
x_{5},,x_{6}\leftrightarrow x_{7},x_{8}\mapsto x_{8},x_{9}\mapsto
x_{9},i\mapsto -i.  \notag
\end{eqnarray}%
A similar computation leads to the dual antilinear symmetry corresponding
for $\omega _{2}=\tau \sigma _{1}\sigma _{3}\sigma _{5}\sigma _{7}$.

Obviously these solutions only capture part of all possibilities as we may
of course also consider the cases $\tilde{h}=4n$ and since (\ref{xcv}) is a
restriction of the most general Ansatz (\ref{Ansatz}). Some solutions
filling these gaps were presented in \cite{FringSmith}. Having been fairly
detailed for the $A_{\ell }$-Weyl group, we will only indicate some selected
examples for reference for the other cases.

\subsection{Antilinearly deformed $B_{\ell }$ root systems}

The simplest class for $\tilde{h}=4$ contains only one element comprised of
two Weyl reflections%
\begin{equation}
\Sigma _{\{1,3,4^{\ell -2}\}}=\left\{ \tilde{\sigma}=\sigma _{\ell -1}\sigma
_{\ell }\right\} .
\end{equation}

\noindent The next class with $\tilde{h}=4$ contains $2\ell -6$ elements 
\begin{equation}
\Sigma _{\{1,2,3,4^{\ell -3}\}}=\left\{ \tilde{\sigma}^{(1,1)},\ldots ,%
\tilde{\sigma}^{(1,\ell -3)},\tilde{\sigma}^{(2,1)},\ldots ,\tilde{\sigma}%
^{(2,\ell -3)}\right\} ,
\end{equation}%
build from a composition of three Weyl reflections{}%
\begin{equation}
\tilde{\sigma}^{(1,i)}:=\sigma _{i}\sigma _{i+2}\sigma _{i+1}\quad \text{%
and\quad }\tilde{\sigma}^{(2,i)}:=\sigma _{\ell }\sigma _{\ell -i-2}\sigma
_{\ell -1}\quad \text{for}\quad i=1,\cdots ,\ell -3.
\end{equation}%
In table 4 we indicate the different types of classes with increasing rank $%
\ell $. We note that whenever the rank increases by one, a new type of class
emerges with one additional Weyl reflection in the element $\tilde{\sigma}$.

{\renewcommand{\tabcolsep}{0.05cm} 
\begin{table}[h]
\begin{tabular}{|c|c|c|c|c|c|c|c|}
\hline
$\ell$ &  &  &  &  &  &  &  \\ \hline
$3$ & $\{1,3,4\}$ &  &  &  &  &  &  \\ \hline
$4$ & $\{1,3,4^2\}$ & $\{1,2,3,4\}$ &  &  &  &  &  \\ \hline
$5$ & $\{1,3,4^3\}$ & $\{1,2,3,4^2\}$ & $\{1,2^2,3,4\}$ &  &  &  &  \\ \hline
$6$ & $\{1,3,4^4\}$ & $\{1,2,3,4^3\}$ & $\{1,2^2,3,4^2\}$ & $\{1^2,2,3^2,4\}$
&  &  &  \\ \hline
$7$ & $\{1,3,4^5\}$ & $\{1,2,3,4^4\}$ & $\{1,2^2,3,4^3\}$ & $%
\{1^2,2,3^2,4^2\}$ & $\{1,2^3,3,4^2\}$ &  &  \\ \hline
$8$ & $\{1,3,4^6\}$ & $\{1,2,3,4^5\}$ & $\{1,2^2,3,4^4\}$ & $%
\{1^2,2,3^2,4^3\}$ & $\{1,2^3,3,4^3\}$ & $\{1^2,2^2,3^2,4^2\}$ &  \\ \hline
$9$ & $\{1,3,4^7\}$ & $\{1,2,3,4^6\}$ & $\{1,2^2,3,4^5\}$ & $%
\{1^2,2,3^2,4^4\}$ & $\{1,2^3,3,4^4\}$ & $\{1^2,2^2,3^2,4^3\}$ &  \\ \hline
$10$ & $\{1,3,4^8\}$ & $\{1,2,3,4^7\}$ & $\{1,2^2,3,4^6\}$ & $%
\{1^2,2,3^2,4^5\}$ & $\{1,2^3,3,4^5\}$ & $\{1^2,2^2,3^2,4^4\}$ &  \\ \hline
\vdots & \vdots & \vdots & \vdots & \vdots & \vdots & \vdots & \ldots \\ 
\hline
$\ell$ & $\{1,3,4^{\ell-2}\}$ & $\{1,2,3,4^{\ell-3}\}$ & $%
\{1,2^2,3,4^{\ell-4}\}$ & $\{1^2,2,3^2,4^{\ell-5}\}$ & $\{1,2^3,3,4^{\ell-5}%
\}$ & $\{1,2^4,3,4^{\ell-6}\}$ & \ldots \\ \hline
\end{tabular}%
\caption{Similarity classes in $B_{\ell }$ with $\tilde{h}=4$.}
\end{table}
} We also report explicitly one deformation matrix resulting from these
classes for an example for which no solution exists with the assumptions
made in \cite{FringSmith}. Using for $B_{5}$ the same general Ansatz as for
the $A_{\ell }$-case in (\ref{xcv}), we obtain for the antilinearly deformed
symmetry of $\tilde{\sigma}^{(1,1)}$ the solution%
\begin{equation}
\theta _{\varepsilon }=\left( 
\begin{array}{ccccc}
\text{$r_{0}$}-i\vartheta & -2i\vartheta & \text{$r_{0}$}-i\vartheta -1 & 0
& 0 \\ 
2i\vartheta & 2\text{$r_{0}$}+2i\vartheta -1 & 2i\vartheta & 0 & 0 \\ 
\text{$r_{0}$}-i\vartheta -1 & -2i\vartheta & \text{$r_{0}$}-i\vartheta & 0
& 0 \\ 
1-\text{$r_{0}$} & 1-\text{$r_{0}$}+i\vartheta & 1-\text{$r_{0}$} & 1 & 0 \\ 
0 & 0 & 0 & 0 & 1%
\end{array}%
\right) ~.  \label{rt}
\end{equation}%
We compute the dual map $\delta ^{\star }$ by solving (\ref{dual}) for the
dual deformation matrix $\theta _{\varepsilon }^{\star }$ with the explicit
form for $\theta _{\varepsilon }$ as in (\ref{rt}). For the standard root
representation of the $B_{\ell }$-roots $(\alpha _{i})_{j}=\delta
_{ij}-\delta _{(i+1)j}$, $(\alpha _{\ell })_{j}=\delta _{\ell j}$, $%
i=1,2,\ldots ,\ell -1$, $j=1,2,\ldots ,\ell $ we obtain%
\begin{equation}
\theta _{\varepsilon }^{\star }=\left( 
\begin{array}{ccccc}
\text{$r_{0}$} & -i\vartheta & i\vartheta & 1-\text{$r_{0}$} & 0 \\ 
i\vartheta & \text{$r_{0}$} & 1-\text{$r_{0}$} & -i\vartheta & 0 \\ 
-i\vartheta & 1-\text{$r_{0}$} & \text{$r_{0}$} & i\vartheta & 0 \\ 
1-\text{$r_{0}$} & i\vartheta & -i\vartheta & \text{$r_{0}$} & 0 \\ 
0 & 0 & 0 & 0 & 1%
\end{array}%
\right) .  \label{B5}
\end{equation}%
By construction the corresponding root space $\tilde{\Delta}^{\star
}(\varepsilon )$ is invariant under the action of some antilinear maps $%
\varpi ^{\star }$, obtained by solving (\ref{xc}). For $\omega _{1}=\tau
\sigma _{1}\sigma _{3}$ we compute the dual antilinear transformation to%
\begin{equation}
\omega _{1}^{\star }=\tau \left( 
\begin{array}{ccccc}
-2i\vartheta (2\text{$r_{0}$}-1) & (1-2\text{$r_{0}$})^{2} & -4(\text{$r_{0}$%
}-1)\text{$r_{0}$} & 2i\vartheta (2\text{$r_{0}$}-1) & 0 \\ 
(1-2\text{$r_{0}$})^{2} & 2i\vartheta (2\text{$r_{0}$}-1) & -2i\vartheta (2%
\text{$r_{0}$}-1) & -4(\text{$r_{0}$}-1)\text{$r_{0}$} & 0 \\ 
-4(\text{$r_{0}$}-1)\text{$r_{0}$} & -2i\vartheta (2\text{$r_{0}$}-1) & 
2i\vartheta (2\text{$r_{0}$}-1) & (1-2\text{$r_{0}$})^{2} & 0 \\ 
2i\vartheta (2\text{$r_{0}$}-1) & -4(\text{$r_{0}$}-1)\text{$r_{0}$} & (1-2%
\text{$r_{0}$})^{2} & -2i\vartheta (2\text{$r_{0}$}-1) & 0 \\ 
0 & 0 & 0 & 0 & 1%
\end{array}%
\right) .  \label{star}
\end{equation}%
The action on the variables amounts with (\ref{star}) simply to 
\begin{eqnarray}
\omega _{1}^{\star }:\tilde{\Delta}^{\ast }(\varepsilon )\rightarrow \tilde{%
\Delta}^{\ast }(\varepsilon ), &&\tilde{x}_{1}\leftrightarrow \tilde{x}_{2},%
\tilde{x}_{3}\leftrightarrow \tilde{x}_{4},\tilde{x}_{5}\mapsto \tilde{x}%
_{5}, \\
&&x_{1}\leftrightarrow x_{2},x_{3}\leftrightarrow x_{4},x_{5}\mapsto
x_{5},i\mapsto -i.
\end{eqnarray}%
A similar computation leads to the dual antilinear symmetry corresponding to 
$\omega _{2}=\tau \sigma _{2}$.

\subsection{Antilinearly deformed $C_{\ell }$ root systems}

A simple class for $\tilde{h}=4$ with only one element $\tilde{\sigma}%
=\sigma _{1}\sigma _{3}\sigma _{2}$ is the case $\Sigma _{\{1,2,3,4^{\ell
-3}\}}$. We present the deformation matrix for the $C_{4}$-case resulting
from this element

\begin{equation}
\theta _{\varepsilon }=\left( 
\begin{array}{cccc}
r_{0}-i\vartheta & -2i\vartheta & r_{0}-i\vartheta -1 & 0 \\ 
2i\vartheta & 2r_{0}+2i\vartheta -1 & 2i\vartheta & 0 \\ 
r_{0}-i\vartheta -1 & -2i\vartheta & r_{0}-i\vartheta & 0 \\ 
2\left( 1-r_{0}\right) & 2\left( 1-r_{0}\right) +2i\vartheta & 2\left(
1-r_{0}\right) & 1%
\end{array}%
\right) .
\end{equation}%
Note that the classes for $C_{\ell }$ are the same as those for $B_{\ell }$,
albeit the deformation matrices are different due to the difference of the
Weyl reflections.

\subsection{Antilinearly deformed $D_{\ell }$ root systems}

In this case a simple class for $\tilde{h}=4$ contains $\ell -1~$elements 
\begin{equation}
\Sigma _{\{1,3,4^{\ell -3}\}}=\left\{ \tilde{\sigma}^{(1)},\tilde{\sigma}%
^{(2)},\ldots ,\tilde{\sigma}^{(\ell =2)},\tilde{\sigma}^{(\ell )}\right\} .
\end{equation}%
with%
\begin{equation}
\tilde{\sigma}^{(i)}=\sigma _{i}\sigma _{i+2}\sigma _{i+1}\quad \text{%
and\quad }\tilde{\sigma}^{(\ell )}=\sigma _{\ell -3}\sigma _{\ell }\sigma
_{\ell -2}\quad \text{for }\quad i=1,\cdots ,\ell -2.
\end{equation}%
As an example for a deformation matrix for $D_{4}$ we present the one
resulting from $\tilde{\sigma}^{(1)}=\sigma _{1}\sigma _{3}\sigma _{2}$

\begin{equation}
\theta _{\varepsilon }=\left( 
\begin{array}{cccc}
r_{0}-i\vartheta & -2i\vartheta & r_{0}-i\vartheta -1 & 0 \\ 
2i\vartheta & 2r_{0}+2i\vartheta -1 & 2i\vartheta & 0 \\ 
r_{0}-i\vartheta -1 & -2i\vartheta & r_{0}-i\vartheta & 0 \\ 
1-r_{0}-i\vartheta & 2-2r_{0} & 1-r_{0}-i\vartheta & 1%
\end{array}%
\right) .
\end{equation}

\subsection{Antilinearly deformed $E_{6+n}$ root systems}

We may treat the exceptional algebras together using for the labeling the $%
E_{8}$-convention in \cite{FringSmith2} and removing vertices from the long
end Dynkin diagram to obtain the $E_{7}$ and $E_{6}$-cases. A simple class
for $\tilde{h}=4$ contains $n+5~$elements 
\begin{equation}
\Sigma _{\{{1,2,3,4^{3}}\}}=\left\{ \sigma _{1}\sigma _{3}\sigma _{4},\sigma
_{1}\sigma _{5}\sigma _{4},\sigma ^{(2)},\sigma ^{(3)}\ldots ,\sigma
^{(n+4)}\right\} ,
\end{equation}%
with $\sigma ^{(i)}=\sigma _{i}\sigma _{i+2}\sigma _{i+1}$ for $i=2,..,n+4.$
The deformation matrix for $\sigma ^{(2)}=\sigma _{3}\sigma _{2}\sigma _{4}$
is computed to 
\begin{equation}
\theta _{\varepsilon }=\left( 
\begin{array}{ccccccc}
1 & 1-r_{0} & -r_{0}-i\vartheta +1 & 1-r_{0} & 0 & 0 & \cdots \\ 
0 & r_{0}+i\vartheta & 2i\vartheta & r_{0}+i\vartheta -1 & 0 & 0 & \cdots \\ 
0 & -2i\vartheta & 2r_{0}-2i\vartheta -1 & -2i\vartheta & 0 & 0 & \cdots \\ 
0 & r_{0}+i\vartheta -1 & 2i\vartheta & r_{0}+i\vartheta & 0 & 0 & \cdots \\ 
0 & 1-r_{0} & -r_{0}-i\vartheta +1 & 1-r_{0} & 1 & 0 & \cdots \\ 
\cdots & \cdots & \cdots & \cdots & \cdots & \cdots &  \\ 
0 & 0 & 0 & 0 & 0 & 1 & 
\end{array}%
\right) .
\end{equation}%
A further class is $\Sigma _{\{1,2^{2},3,4^{2+n}\}}$ with elements $\tilde{%
\sigma}=\sigma _{1}\sigma _{4}\sigma _{2}\sigma _{5},\ldots $

\subsection{Antilinearly deformed $F_{4}$ root systems}

The simplest class for $\tilde{h}=4$ contains only one element%
\begin{equation}
\Sigma _{\{1,3,4^{2}\}}=\left\{ \sigma _{3}\sigma _{2}\right\} .
\end{equation}%
The deformation matrix is computed to%
\begin{equation}
\theta _{\varepsilon }=\left( 
\begin{array}{cccc}
1 & 2\left( 1-r_{0}\right) & 2\left( 1-r_{0}\right) -2i\vartheta & 0 \\ 
0 & 2r_{0}+2i\vartheta -1 & 4i\vartheta & 0 \\ 
0 & -2i\vartheta & 2r_{0}-2i\vartheta -1 & 0 \\ 
0 & 1-r_{0}+i\vartheta & 2\left( 1-r_{0}\right) & 1%
\end{array}%
\right) .
\end{equation}

\section{Deformation matrices from two arbitrary elements in $\mathcal{W}$}

The procedure outlined in section 2 is entirely generic and may of course
also be carried out by starting from any arbitrary elements in $\mathcal{W}$
\ different from $\sigma _{+}$ and $\sigma _{-}$. Due to the random choice
we allow for the symmetries we have to consider now concrete cases. It is
instructive to discuss some examples for which no nontrivial solutions were
found previously.

Let us therefore consider $B_{3}$. As an abstract Coxeter group $B_{3}$ is
fully characterized by three involutory generators $\sigma _{1}^{2}=\sigma
_{2}^{2}=\sigma _{3}^{2}=\mathbb{I}$ together with the three relations $%
\sigma _{1}\sigma _{3}=\sigma _{3}\sigma _{1}$, $\sigma _{1}\sigma
_{2}\sigma _{1}=\sigma _{2}\sigma _{1}\sigma _{2}$ and $\sigma _{2}\sigma
_{3}\sigma _{2}\sigma _{3}=\sigma _{3}\sigma _{2}\sigma _{3}\sigma _{2}$.
Choosing now in (\ref{o1}) the involutions different from the previous
section as $\hat{\omega}_{1}=\sigma _{1}$ and $\hat{\omega}_{2}=\sigma
_{1}\sigma _{3}$ yields $\Omega _{12}=\sigma _{3}$. Thus we have taken $\hat{%
\omega}_{1}$ and $\hat{\omega}_{2}$ both to be factors in $\sigma _{-}$.
According to (\ref{exp}) we have to identify next all elements in $B_{3}$
commuting with $\sigma _{3}$. Using the three relations and the three
generators we find $\{\mathbb{I},\sigma _{1},\sigma _{3},\sigma _{1}\sigma
_{3},\sigma _{1}\sigma _{2}\sigma _{3}\sigma _{2}\}$ leading to the Ansatz%
\begin{equation}
\theta _{\varepsilon }=r_{0}(\varepsilon )\mathbb{I+}r_{1}(\varepsilon
)\sigma _{1}+r_{2}(\varepsilon )\sigma _{3}+r_{3}(\varepsilon )\sigma
_{1}\sigma _{3}+r_{4}(\varepsilon )\sigma _{1}\sigma _{2}\sigma _{3}\sigma
_{2}.  \label{tt}
\end{equation}%
which solves all the constraints (\ref{const}) when%
\begin{equation}
r_{0}=\sqrt{1-r_{4}^{2}}-r_{2},\quad r_{1}=-r_{3}=1+r_{2}-\frac{r_{4}}{2}-%
\frac{1}{2}\sqrt{1-r_{4}^{2}},\quad \lim\limits_{\varepsilon \rightarrow
0}r_{4}=0,\quad r_{4}\in i\mathbb{R}.
\end{equation}%
Upon substitution into the Ansatz (\ref{tt}) the deformation matrix takes on
the form%
\begin{equation}
\theta _{\varepsilon }=\left( 
\begin{array}{ccc}
\sqrt{1-r_{4}^{2}}+r_{4} & 2r_{4} & 2r_{4} \\ 
-r_{4} & ~~\sqrt{1-r_{4}^{2}}-r_{4} & ~~\sqrt{1-r_{4}^{2}}-r_{4}-1 \\ 
0 & 0 & 1%
\end{array}%
\right) .  \label{ty}
\end{equation}%
Thus we have only one free function left. The same Ansatz (\ref{tt}) can be
used for the choice $\hat{\omega}_{1}=\sigma _{2}$ and $\hat{\omega}%
_{2}=\sigma _{2}\sigma _{3}$, but in that case we are led to the trivial
solution $\theta _{\varepsilon }=\mathbb{I}$.

With the help of (\ref{ty}) we may now also find the dual map $\delta
^{\star }$ by solving (\ref{dual}) for the dual deformation matrix $\theta
_{\varepsilon }^{\star }$. For the standard root representation of $B_{3}$
we obtain

\begin{equation}
\theta _{\varepsilon }^{\star }=\left( 
\begin{array}{lll}
\sqrt{1-r_{4}^{2}} & -r_{4} & 0 \\ 
r_{4} & \sqrt{1-r_{4}^{2}} & 0 \\ 
0 & 0 & 1%
\end{array}%
\right) .
\end{equation}%
The corresponding root space $\tilde{\Delta}^{\star }(\varepsilon )$ is then
by construction invariant under the action of some antilinear maps $\varpi
^{\star }$, obtained by solving (\ref{xc}). For $\omega _{1}=\tau \sigma
_{1} $ and $\omega _{2}=\tau \sigma _{1}\sigma _{3}$ we compute%
\begin{eqnarray}
\omega _{1}^{\star } &:&\tilde{\Delta}^{\ast }(\varepsilon )\rightarrow 
\tilde{\Delta}^{\ast }(\varepsilon ),~~\tilde{x}_{1}\leftrightarrow \tilde{x}%
_{2},\tilde{x}_{3}\mapsto \tilde{x}_{3}\equiv x_{1}\leftrightarrow
x_{2},x_{3}\mapsto x_{3},i\mapsto -i, \\
\omega _{2}^{\star } &:&\tilde{\Delta}^{\ast }(\varepsilon )\rightarrow 
\tilde{\Delta}^{\ast }(\varepsilon ),~~~\tilde{x}_{1}\leftrightarrow \tilde{x%
}_{2},\tilde{x}_{3}\mapsto -\tilde{x}_{3}\equiv x_{1}\leftrightarrow
x_{2},x_{3}\mapsto -x_{3},i\mapsto -i.
\end{eqnarray}%
Taking $r_{4}=\pm i\sinh \varepsilon $ we notice that $\theta _{\varepsilon
}^{\star }$ becomes a rotation about the complex angle $\pm i\varepsilon $
for the variables $x_{1}$ and $x_{2}$ accompanied by a reflection in $x_{3}$
for the latter case.

\section{Deformation matrices from rotations in the dual space}

So far we have started with given antilinear involution $\varpi _{i}$ and
constructed the deformation map $\delta $ by solving the constraints (\ref%
{const}) for a given Weyl group, i.e.~given some $\omega _{i}$ we determined
the deformation matrix $\theta _{\varepsilon }$. Subsequently we constructed
the corresponding maps $\delta ^{\star }$ and $\varpi ^{\star }$acting in
the dual spaces. We may also try to reverse the procedure and start with the
dual space with given maps $\delta ^{\star }$ and $\varpi ^{\star }$ and
determine the maps $\varpi _{i}$ and $\delta $ thereafter. In view of the
last section it is natural to assume the $\theta _{\varepsilon }^{\star }$
to be an element of the special orthogonal group. We define therefore the $%
(2n+1)\times (2n+1)$-matrix%
\begin{equation}
\theta _{\varepsilon }^{\star }=\left( 
\begin{array}{ccccc}
R &  &  &  &  \\ 
& R &  & 0 &  \\ 
&  & R &  &  \\ 
& 0 &  & \ddots &  \\ 
&  &  &  & 1%
\end{array}%
\right) \qquad \text{with }R=\left( 
\begin{array}{rr}
\cosh \varepsilon & i\sinh \varepsilon \\ 
-i\sinh \varepsilon & \cosh \varepsilon%
\end{array}%
\right) ,  \label{re}
\end{equation}%
and construct the deformation matrix $\theta _{\varepsilon }$ by solving (%
\ref{dual}). We note that this constraint might not admit any solutions for
certain Weyl groups. In fact for the standard representation for $A_{\ell }$
it is easy to verify that indeed there exists no solution. However, for the
special orthogonal Weyl groups $B_{\ell }\equiv SO(2\ell +1)$ and $D_{\ell
}\equiv SO(2\ell )$ one can solve (\ref{dual}). Since previously we did not
find solutions for odd ranks in the $B$-series in \cite{FringSmith} based on
the assumptions made in there, we present here some solutions for $B_{2n+1}$%
. Solving (\ref{xc}) for $\theta _{\varepsilon }$ using the standard
representation for the $B_{\ell }$-roots we compute the deformed roots to%
\begin{eqnarray}
\tilde{\alpha}_{2j-1} &=&\cosh \varepsilon \alpha _{2j-1}+i\sinh \varepsilon
\left( \alpha _{2j-1}+2\sum\limits_{k=2j}^{\ell }\alpha _{k}\right) ~~~~~%
\text{for }j=1,\ldots ,n, \\
\tilde{\alpha}_{2j} &=&\cosh \varepsilon \alpha _{2j}-i\sinh \varepsilon
\left( \sum\limits_{k=2j}^{2j+2}\alpha _{k}+2\sum\limits_{k=2j+3}^{\ell
}2\alpha _{k}\right) ~~\text{for }j=1,\ldots ,n-1, \\
\tilde{\alpha}_{\ell -1} &=&\cosh \varepsilon (\alpha _{\ell -1}+\alpha
_{\ell })-\alpha _{\ell }-i\sinh \varepsilon \left( \alpha _{\ell -2}+\alpha
_{\ell -1}+\alpha _{\ell }\right) , \\
\tilde{\alpha}_{\ell } &=&\alpha _{\ell }.
\end{eqnarray}%
By construction we have satisfied the last three constraints in (\ref{const}%
). Furthermore, we find that $\theta _{\varepsilon }^{\ast }\sigma
_{-}=\sigma _{-}\theta _{\varepsilon }$ but $\theta _{\varepsilon }^{\ast
}\sigma _{+}\neq \sigma _{+}\theta _{\varepsilon }$ with $\sigma
_{-}=\prod\nolimits_{k=1}^{n+1}$ $\sigma _{2k-1}$ and $\sigma
_{+}=\prod\nolimits_{k=1}^{n}$ $\sigma _{2k}$. Thus in this case $\tau
\sigma _{+}$ does not constitute an antilinear symmetry which implies that $%
\left[ \sigma ,\theta _{\varepsilon }\right] \neq 0$. This is the reason why
this solution has escaped the previous analysis. However, besides under the
action of $\omega _{-}:=\tau \sigma _{-}=\sigma _{-}^{\varepsilon }$ the
root space $\tilde{\Delta}(\varepsilon )$ is left invariant under various
other antilinear maps which consist of subfactors of $\sigma _{-}$. For $%
B_{3}$ we observed this in section 4 with $\sigma _{-}=\sigma _{1}\sigma
_{3} $ and $\sigma _{3}$ being the additional symmetry. A generalization to $%
B_{2n}$ is straightforward simply by starting in (\ref{re}) with an $%
(2n)\times (2n)$-matrix of the form (\ref{re}) without the entry $1$.

Similarly as for $B_{2n+1}$ we may also solve (\ref{dual}) for the $D_{2n}$
Weyl group for which we demonstrated in \cite{FringSmith} that no solution
to the constraining equations (\ref{const}) based on the Ansatz (\ref{th})
could exist, that is for given invariance $\sigma _{-}^{\varepsilon }$ and $%
\sigma _{+}^{\varepsilon }$ . Starting with $\theta _{\varepsilon }^{\star }$
in the form (\ref{re}) we construct the deformed roots with standard
representation for the $D_{\ell }$-roots $(\alpha _{i})_{j}=\delta
_{ij}-\delta _{(i+1)j}$, $(\alpha _{\ell })_{j}=\delta _{j(\ell -1)}+\delta
_{j\ell }$, $i=1,2,\ldots ,\ell -1$, $j=1,2,\ldots ,\ell $ as 
\begin{eqnarray}
\tilde{\alpha}_{\ell -(2j+1)} &=&\cosh \varepsilon \alpha _{\ell
-(2j+1)}+i\sinh \varepsilon \left[ \sum\limits_{k=\ell -(2j+1)}^{\ell
}\alpha _{k}+\sum\limits_{\ell -2j}^{\ell -2}\alpha _{k}\right] , \\
\tilde{\alpha}_{\ell -(2j+2)} &=&\cosh \varepsilon \alpha _{\ell
-(2j+2)}-i\sinh \varepsilon \left[ \sum\limits_{k=\ell -2j-3}^{\ell }\alpha
_{k}+\sum\limits_{\ell -2j}^{\ell -2}\alpha _{k}\right] , \\
\tilde{\alpha}_{\ell -2} &=&\cosh \varepsilon \alpha _{\ell -2}-i\sinh
\varepsilon (\alpha _{\ell -3}+\alpha _{\ell -2}+\alpha _{\ell }), \\
\tilde{\alpha}_{\ell -1} &=&\cosh \varepsilon \alpha _{\ell -1}+i\sinh
\varepsilon \alpha _{\ell }, \\
\tilde{\alpha}_{\ell } &=&\cosh \varepsilon \alpha _{\ell }-i\sinh
\varepsilon \alpha _{\ell -1}.
\end{eqnarray}

Similarly as for the $B_{2n+1}$ case we find that $\theta _{\varepsilon
}^{\ast }\sigma _{-}=\sigma _{-}\theta $ whereas $\theta _{\varepsilon
}^{\ast }\sigma _{+}\neq \sigma _{+}\theta $ with $\sigma
_{-}=\prod_{k=1}^{n}\sigma _{2k-1}$ and $\sigma _{+}=\prod_{k=1}^{n-2}\sigma
_{2k}$. Again it is easy enough to generalize this to the the $D_{2n+1}$
case.

For the standard $(n+1)$-dimensional representation of $A_{\ell }$ a
rotation on a subspace of $\tilde{\Delta}^{\ast }(\varepsilon )$ for the
first two coordinates and its conjugate momenta was suggested in \cite%
{Ghosh1,Ghosh2}. In that case, and for its generalisation (\ref{re}), the
corresponding deformation $\tilde{\Delta}(\varepsilon )$ can not be
constructed since (\ref{dual}) admits no solution.

\section{The construction of q-deformed roots}

Mainly motivated by an applications to affine Toda field theories in mind we
provide in this section a construction for $q$-deformed roots, meaning that
we are seeking a map 
\begin{equation}
\delta _{q}:~\Delta \subset \mathbb{R}^{n}\rightarrow \Delta _{q}\subset 
\mathbb{R}^{n}[q],\qquad \alpha \mapsto \alpha _{q}=\Theta _{q}\alpha ,
\end{equation}%
with $\mathbb{R}^{n}[q]$ denoting a polynomial ring in $q\in \mathbb{C}$. In
this case the complex deformation matrix $\Theta _{q}$ depends on the
deformation $q$ in such a way that $\lim_{q\rightarrow 1}\Theta _{q}=\mathbb{%
I}$. Our construction is centered around a q-deformation of the Coxeter
element in the factorised form already used in this manuscript $\sigma
:=\sigma _{-}\sigma _{+}$ as introduced in \cite{q1,q2} 
\begin{equation}
\sigma _{q}:=\sigma _{-}^{q}\,\tau _{q}\,\sigma _{+}^{q}\,\tau _{q}\,\,.
\label{sq}
\end{equation}%
The deformations of the Coxeter factors $\sigma _{\pm }$ are defined as%
\begin{equation}
\sigma _{\pm }^{q}:=\prod\limits_{i\in V_{\pm }}\sigma _{i}^{q}\,\,,
\end{equation}%
where the product is taken over $q$-deformed Weyl reflections, whose action
on simple roots $\alpha _{i}\in \Delta $ is given as 
\begin{equation}
\sigma _{i}^{q}(\alpha _{j}):=\alpha _{j}-(2\delta _{ij}-\left[ I_{ji}\right]
_{q})\alpha _{i}\,\,.
\end{equation}%
We employed here one of the standard definition for a $q$-deformed integer%
\footnote{%
We will frequently use the identities $[1]_{q}=1$, $[2]_{q}=q+q^{-1}$ and $%
[3]_{q}=1+q^{2}+q^{-2}$.} 
\begin{equation}
\lbrack n]_{q}:=\frac{q^{n}-q^{-n}}{q-q^{-1}}\,.  \label{qqq}
\end{equation}%
A further deformation in $q$ results from the map $\tau _{q}$ also employed
in (\ref{sq}) 
\begin{equation}
\tau _{q}(\alpha _{i}):=q^{t_{i}}\alpha _{i}\,\,.  \label{tau}
\end{equation}%
The integers $t_{i}$ are the symmetrizers of the incidence matrix $I$, i.e.~$%
I_{ij}t_{j}=I_{ji}t_{i}$. From these definitions it is evident the
q-deformed Coxeter element is only different from the ordinary one when the
associated Weyl group is related to non-simply laced algebras.

Since $\sigma _{q}$ is defined by its action on the simple roots $\alpha $
it is natural to seek an operator $\mathcal{O}_{q}$ acting on elements $%
\alpha _{q}\in \Delta _{q}$ with the appropriate limit $\lim_{q\rightarrow 1}%
\mathcal{O}_{q}=\mathbb{\sigma }$. Recalling that the order of $\sigma $ is
the Coxeter number $h$, i.e.~$\sigma ^{h}=1$, whereas the order of $\sigma
_{q}$ is deformed $\sigma _{q}^{h}=q^{2H}$, it is obvious that the relation
can not be a simple similarity transformation. Here $H$ is the $\ell $-th
Coxeter number of the dual algebra, see e.g.~\cite{Kac} for more details.
Therefore we make the Ansatz 
\begin{equation}
\sigma _{q}\alpha =q^{2H/h}\Theta _{q}^{-1}\sigma \Theta _{q}\,\alpha
\,=q^{2H/h}\Theta _{q}^{-1}\sigma \,\alpha _{q}.  \label{sim2}
\end{equation}%
and readily identify the operator $\mathcal{O}_{q}=$ $q^{2H/h}\Theta
_{q}^{-1}\sigma $. The relation (\ref{sim2}) serves as the defining relation
for the $q$-deformed simple roots $\alpha _{q}=\Theta _{q}\,\alpha $.

In analogy to the undeformed situation we introduce a $q$-deformed simple
root dressed by a colour value as a separate quantity $\left( \gamma
_{q}\right) _{i}:=c_{i}\left( \alpha _{q}\right) _{i}$. This serves as a
representant to introduce the $q$-deformed Coxeter orbits 
\begin{equation}
\left( \Omega _{q}\right) _{i}:=\left\{ \left( \gamma _{q}\right)
_{i},\sigma \left( \gamma _{q}\right) _{i},\ldots ,\sigma ^{h-1}\left(
\gamma _{q}\right) _{i}\right\} \,.  \label{qorb}
\end{equation}%
The entire $q$-deformed root system $\Delta _{q}$ is then spanned by the
union of all $\ell $ $q$-deformed Coxeter orbits 
\begin{equation}
\Delta _{q}:=\bigcup_{i=1}^{\ell }\left( \Omega _{q}\right) _{i}\,\,.
\end{equation}%
At this stage it is not obvious under which type of symmetry\ $\Delta _{q}$
is left invariant.

\subsection{The q-deformed root space for $\left(
C_{2}^{(1)},D_{3}^{(2)}\right) $}

Let is now illustrate the working of the above formulae with a simple
explicit example. The incidence matrix for $C_{2}$ is in this case defined
as $I_{12}=1$, $I_{21}=2$, such that the symmetrizers are $t_{1}=1$ and $%
t_{2}=2$. The Coxeter numbers are $h=4$ and $H=6$. Therefore we obtain%
\begin{equation}
\sigma _{-}^{q}=\left( 
\begin{array}{cc}
-1 & 0 \\ 
\lbrack 2]_{q} & 1%
\end{array}%
\right) ,\quad \sigma _{+}^{q}=\left( 
\begin{array}{cc}
1 & 1 \\ 
0 & -1%
\end{array}%
\right) ,\quad \tau _{q}=\left( 
\begin{array}{cc}
q & 1 \\ 
0 & q^{2}%
\end{array}%
\right) ,\quad \sigma _{q}=q^{2}\left( 
\begin{array}{cc}
-1 & -q \\ 
\lbrack 2]_{q} & 1%
\end{array}%
\right) .
\end{equation}%
Solving equation (\ref{sim2}) then yields the deformed roots%
\begin{eqnarray}
\left( \alpha _{q}\right) _{1} &=&r_{1}\alpha _{1}+\frac{q}{1+q}%
(r_{1}-r_{2})\alpha _{2},  \label{qr1} \\
\left( \alpha _{q}\right) _{2} &=&\frac{r_{2}+(r_{2}-2r_{1})q^{2}}{q+q^{2}}%
\alpha _{1}+r_{2}\alpha _{2},  \label{qr2}
\end{eqnarray}%
where $r_{1}$, $r_{2}$ depend on $q$ with the limiting behaviour $%
\lim_{q\rightarrow 1}r_{1}=1$ and $\lim_{q\rightarrow 1}r_{2}=1$.

\section{Non-Hermitian Calogero models}

We can now formulate and investigate models on these complex root spaces.
Thus we may consider new types of non-Hermitian generalisations of Calogero
models 
\begin{equation}
\mathcal{H}_{0,\varepsilon ,q}(p,x)=\frac{p^{2}}{2}+\frac{\omega ^{2}}{4}%
\sum_{\alpha }(\alpha \cdot x)^{2}+\sum_{\alpha }\frac{g_{\alpha }}{(\alpha
\cdot x)^{2}},\qquad \alpha _{i}\in \Delta ,\tilde{\Delta}(\varepsilon
),\Delta _{q},  \label{HC}
\end{equation}%
or the analogues of Calogero-Moser-Sutherland models when replacing the
rational potential by a trigonometric or elliptic one. The model $\mathcal{H}%
_{\varepsilon }$ for the rational potential has been investigated previously 
\cite{Milos,FZ,FringSmith,FringSmith2} and was found to have remarkable
properties when compared with the standard undeformed models $\mathcal{H}%
_{0} $. As a result of the deformation into the complex domain the
singularities in the potential are regularized. Therefore the models no
longer have to be defined in separate disjointed regimes and continued by
phase factors corresponding to some selected statistics. As was shown in 
\cite{FringSmith}, in the $\mathcal{H}_{\varepsilon }$-models the anyonic
phase factors are automatically present and the models can be defined on the
entire domain of the configuration space. As a consequence the energy
spectra of these models will also be different. Various ground state
wavefunctions and those corresponding to exited states were computed in \cite%
{FringSmith} and \cite{FZ}, respectively.

Since the Hamiltonians $\mathcal{H}_{\varepsilon ,q}$ are not Hermitian the
canonical variables $p$ and $x$ are non-observable in the standard Hilbert
space. However, it is by now well understood how to reconcile this by
constructing a well defined metric operator $\rho $ \cite%
{Bender:2004sa,Mist,JM,CA,KBZ,Mostdel,MGH,PEGA,Bender:2008uu,PEGA2,OlaAnd}.
One seeks a linear, invertible, Hermitian and positive operator acting in
the Hilbert space, such that $\mathcal{H}_{\varepsilon ,q}$ becomes a
self-adjoint operator with regard to this metric such that $p$ and $x$
become observable in this space. For this purpose one constructs a so-called
Dyson map $\eta $, which maps the non-Hermitian Hamiltonian $H$ adjointly to
a Hermitian Hamiltonian $h$ 
\begin{equation}
h=\eta H\eta ^{-1}=h^{\dagger }=(\eta ^{-1})^{\dagger }H^{\dagger }\eta
^{\dagger }~~\Leftrightarrow ~~H^{\dagger }\rho =\rho H\text{ \ with }\rho
=\eta ^{\dagger }\eta \text{.}  \label{1}
\end{equation}%
Depending on the assumptions made on the metric such type of Hamiltonians
are referred to with different terminology. When no assumption is made on
the positivity of the $\rho $ in (\ref{1}), the relation on the right hand
side constitutes the \emph{pseudo-Hermiticity} condition, see e.g.~\cite%
{pseudo1,pseudo2,Mostafazadeh:2001nr}, whenever the operator $\rho $ is
linear, invertible and Hermitian. In case the operator $\rho $ is positive
but not invertible this condition is usually referred to as \emph{%
quasi-Hermiticity} \cite{Dieu,Urubu}. Different terminology is used at times
with a less clear meaning.

In general we can not map the Hamiltonians $\mathcal{H}_{\varepsilon ,q}$ to
some Hermitian counterparts in a very obvious way, but in some case we can
provide the explicit transformation $\eta $. We recall that the rotations in
(\ref{re}) on two variables can be realised by means of the angular momentum
operators $L_{ij}=x_{i}p_{j}-x_{j}p_{i}$ 
\begin{equation}
\binom{\tilde{z}_{i}}{\tilde{z}_{j}}=R_{ij}\binom{z_{i}}{z_{j}}=\eta _{ij}%
\binom{z_{i}}{z_{j}}\eta _{ij}^{-1},\quad \quad \text{for }z\in \{x,p\}\text{%
, }\eta _{ij}=e^{\varepsilon (x_{i}p_{j}-x_{j}p_{i})}.
\end{equation}%
Noting furthermore that 
\begin{equation}
\mathcal{H}_{0}(\tilde{p},\tilde{x})=\mathcal{H}_{\varepsilon }(p,x),
\end{equation}%
we can find many explicit transformations of the type (\ref{1}), which map
these Hamiltonians to some isospectral Hermitian counterpart%
\begin{equation}
\mathcal{H}_{0}(p,x)=\eta \mathcal{H}_{\varepsilon }(p,x)\eta ^{-1}.
\end{equation}%
For instance for the $B_{\ell }$-models based on the deformations (\ref{re})
the Dyson map is simply%
\begin{equation}
\eta =\eta _{12}^{-1}\eta _{34}^{-1}\eta _{56}^{-1}\ldots \eta _{(\ell
-2)(\ell -1)}^{-1}.
\end{equation}%
In other cases based on special orthogonal groups the rotations involved
might not commute. For instance, for the $B_{5}$-model based on the
deformation (\ref{B5}) with $r_{0}=\cosh ^{2}\varepsilon $ we find that%
\begin{equation}
\tilde{x}=\theta _{\varepsilon }^{\star
}x=R_{24}^{-1}R_{13}R_{34}R_{12}^{-1}x=\eta x\eta ^{-1}\text{,\qquad with }%
\eta =\eta _{24}^{-1}\eta _{13}\eta _{34}\eta _{12}^{-1}.
\end{equation}%
When the deformation in the configuration space is not based on rotations
such that inner products are not preserved it remains a challenge to find
the corresponding Dyson maps and isospectral Hermitian couterparts. We also
leave the investigation for the $\mathcal{H}_{q}(p,x)$-models for further
investigations.

\section{Non-Hermitian affine Toda theories}

One of the main obstacles to overcome when passing from a classical
description of a field theory to a full-fledged quantum field theory is
renormalisation. In 1+1 space-time dimensions many miracles occur which
allow to express a number of physical quantities in an exact, that is
non-perturbative, manner. In particular it is possible to formulate
classical Lagrangians which are in some sense exact from the quantum field
theory point of view. The classical affine Toda field theory is a prototype
for this kind of behaviour and has the remarkable property that its the
classical mass ratios remain preserved in the quantum field theory after
renormalisation, whenever the associated Lie algebra is simply laced \cite%
{Arin,BCDS,DDV,CM1,PD1,PD2,Mass2,Mussardo:1992uc}. This property ceases to
be valid when the algebra becomes non-simply laced \cite%
{Del3,KW,WW,CDS,PD,Khast,q1,q2}, in which case one has to consider a dual
pair of affine Lie algebras \cite{Kac} and the quantum mass ratios
interpolate via an effective coupling constant between the values obtained
from these two algebras. In the strong and weak limit of the coupling
constant either of these two cases is obtained.

One may now pose the question whether it is also possible to formulate some
naturally modified Lagrangians for non-simply laced algebras which already
capture some exact features from the quantum level, such as preserving the
classical mass ratios when renormalised. In addition we may study models in
which the roots are elements of the antilinearly invariant space. In terms
of simple roots we consider now the three different versions of affine Toda
field theories defined by the Lagrangians 
\begin{equation}
\mathcal{L}_{0,\varepsilon ,q}:=\frac{1}{2}\sum\limits_{i=1}^{\ell }\partial
_{\mu }\phi _{i}\partial ^{\mu }\phi _{i}-\frac{m^{2}}{\beta ^{2}}%
\sum\limits_{i=0}^{\ell }n_{i}e^{\beta \alpha _{i}\cdot \phi },\qquad \alpha
_{i}\in \Delta ,\tilde{\Delta}(\varepsilon ),\Delta _{q}.\,\,  \label{L}
\end{equation}%
The Lagrangian $\mathcal{L}_{0}$ corresponds to the standard version whereas 
$\mathcal{L}_{\varepsilon ,q}$ are newly proposed models. The $\ell $
components of $\phi $ are real scalar fields, $m$ an overall mass scale and
the $\beta $ is the coupling constant. The $\alpha $'s are simple roots with 
$\alpha _{0}$ being the negative of the longest root, whose expansion in
terms of simple roots in the relevant spaces $\alpha
_{0}=-\sum\nolimits_{i=1}^{\ell }n_{i}\alpha _{i}$ is the defining relation
for the integers $n_{i}$, often referred to as Kac labels. The $\mathcal{L}%
_{0}$ theories are known to fall roughly into two different classes
characterised by $\beta $ taken to be either real or or purely complex in
which case the Yang-Baxter equation obeyed by the scattering matrix is
either trivial or non-trivial, respectively. When $\beta \in i\mathbb{R}$
the theory is in general non-Hermitian, except for the $A_{2}$-case
corresponding to the sine-Gordon model, but the classical mass spectra were
still found to be real and stable with respect to small perturbations \cite%
{Holl}. Here we conjecture that the $\mathcal{L}_{\varepsilon ,q}$-models
are also meaningful.

The classical mass matrix for the scalar fields is simply given by the
quadratic term in the fields of the Lagrangian and is easily extracted from
the formulation (\ref{L}) 
\begin{equation}
M_{ij}^{2}=m^{2}\sum\limits_{a=0}^{\ell }n_{a}\alpha _{a}^{i}\alpha
_{a}^{j}\,,\qquad \alpha _{i}\in \Delta ,\tilde{\Delta}(\varepsilon ),\Delta
_{q}.  \label{massc}
\end{equation}%
The mathematical fact that the overall length of the roots is a matter of
convention is reflected in the physical property that the overall mass scale
is not fixed. This is captured in the constant $m$.

\subsection{The mass spectrum of $\left( C_{2}^{(1)},D_{3}^{(2)}\right) $-$%
\mathcal{L}_{q}$}

Taking the two $q$-deformed simple roots to be of the form (\ref{qr1}), (\ref%
{qr2}), noting that the Kac labels for $C_{2}$ are $n_{1}=2$, $n_{2}=1$ and
using the non-standard representation for the undeformed\ $C_{2}$-roots $%
\alpha _{1}=\{0,1\}$, $\alpha _{2}=\{1,-1\}$ we compute the mass matrix in (%
\ref{massc}). The virtue of this basis is that in the limit $q\rightarrow 1$
the mass matrix is diagonal. For $q\neq 1$ the direct evaluation leads to a
nondiagonal matrix. However, imposing the additional constraint%
\begin{equation}
r_{2}=r_{1}q\frac{3q^{2}-5q+2+(q+1)\sqrt{(16-7q)q-8}}{2\left(
2q^{3}-q^{2}+q-1\right) },
\end{equation}%
eliminates the off-diagonal elements. We obtain%
\begin{eqnarray}
M_{11}^{2} &=&r_{1}^{2}q^{3}\frac{2q^{3}+8q^{2}-7q+\left( 1-2q^{2}\right) 
\sqrt{16q-7q^{2}-8}}{\left( 1-2q^{3}+q^{2}-q\right) ^{2}},  \label{M11} \\
M_{22}^{2} &=&r_{1}^{2}q\frac{11q^{5}-18q^{4}+19q^{3}-10q^{2}+q+\left(
q^{4}+2q^{3}-3q^{2}+2q-1\right) \sqrt{16q-7q^{2}-8}}{\left(
2q^{3}-q^{2}+q-1\right) ^{2}},  \notag  \label{M22}
\end{eqnarray}%
with $m_{1}=M_{11}$, $m_{2}=M_{22}$ being the classical masses of the two
scalar fields. As can be found in the above mentioned literature, the
quantum mass ratios of the $\mathcal{L}_{0}$-theory are given by%
\begin{equation}
\frac{m_{1}}{m_{2}}=\frac{\sin \left[ \frac{1}{24}(6-B)\pi \right] }{\cos
\left( \frac{B\pi }{12}\right) },\qquad \text{with }B=\frac{2H\beta ^{2}}{%
H\beta ^{2}+4\pi \ell h},  \label{M1M2}
\end{equation}%
where $B\in \lbrack 0,2]$ denotes the effective coupling constant. From (\ref%
{M11}), (\ref{M22}) and (\ref{M1M2}) we can therefore fix the deformation
parameter such that the quantum mass ratios of $\mathcal{L}_{0}$ correspond
to the classical mass ratios of $\mathcal{L}_{q}$. We find%
\begin{eqnarray}
q &=&\frac{1}{1+\sqrt{3\left( \cos \frac{B\pi }{24}+\sin \frac{B\pi }{24}%
\right) +2\sin \frac{B\pi }{12}-3}}, \\
&=&1-\frac{1}{2}\sqrt{\frac{7\pi }{6}}\sqrt{B}+\frac{7\pi B}{24}-\frac{%
193\pi ^{3/2}B^{3/2}}{192\sqrt{42}}+\frac{95\pi ^{2}B^{2}}{1152}+O\left(
B^{5/2}\right) .
\end{eqnarray}%
Notice that deformation parameter $q(B)$ is a decreasing real valued
function of $B$ taking values between $1$ and $\approx 0.435936$.
Consequently the coefficients in (\ref{qr1}), (\ref{qr2}) in front of the
simple roots acquire complex when the effective coupling constant varies
between $0$ and $2$.

The classical mass spectrum of $\mathcal{L}_{q}$ equals the quantum mass
spectrum of $\mathcal{L}_{0}$.

\section{Conclusions}

We have provided two alternative general methods of construction for complex
root systems. The first is based on using some selected elements of the Weyl
group as analogues of the parity transformation, which are then extended
such that the entire root space remains invariant under at least one
antilinear symmetry. We have provided explicit solutions of different types
for a large number of specific Weyl groups. Since the suggested method is
very generic, i.e.~allowing to start from any element in the Weyl group, it
is useful to select a further principle providing some guidance. Starting
from the factors of the Coxeter element serves for that purpose, but we have
also seen that this is often too restrictive and for certain algebras it
could be shown that no solutions exist in such a setting. However, we
demonstrated that this can be overcome when starting from reduced versions
of these factor. The drawback is then that this gives rise to a large number
of possibilities. Nonetheless, as we demonstrated many of them lie in the
same similarity class, which provides a certain ordering principle. When
giving up even this guiding principle one can still find interesting
solutions. The construction becomes even less restrictive if we also give up
the demand of preserving the inner products. We have paid particular
attention to the construction of the deformed variables in the dual space
together with the corresponding antilinear symmetries. The second type of
construction is based on deformations of the standard Coxeter element. The
complex roots resulting from this procedure are not naturally invariant
under an obvious symmetry.

For the deformations related to the special orthogonal groups we identified
in some cases the corresponding rotations in the dual space. We also
reversed the construction in some examples and identified the corresponding
deformed roots when starting from certain rotations. It would be interesting
to have a precise one-to-one relation between the deformed roots and
deformed variables. We leave this as an open challenge.

Both constructions may be employed in the context of multi-particle systems.
Here we indicated that all non-Hermitian Calogero-Moser-Sutherland models of
the type $\mathcal{H}_{\varepsilon }(p,x)$ based on $B_{\ell }$ and $D_{\ell
}$ Weyl groups may be mapped onto a Hermitian model via similarity
transformations involving various combinations of the angular momentum
operators. For the models based on other Weyl groups we expect this
transformation to exist, but leave the explicit construction for future
investigations. Further interesting open questions for future investigations
are to find the explicit solutions including their modified spectra and to
settle the questions of whether the deformed models are still integrable.

The second type of construction was employed explicitly to define a new type
of non-Hermitian affine Toda theory. These models were found to have the
interesting property that their classical mass ratios are identical in all
orders of the coupling constant to the quantized and renomalised version of
their undeformed couterparts. We leave the interesting problem of
investigating more examples for different types of algebras for the future.

\bigskip

\noindent \textbf{Acknowledgments:} MS is supported by EPSRC.

\newpage

\appendix

\section{Appendix}

In this appendix we provide more examples of reduced root spaces generated
from different types of classes. We exhibit also the action of $\tilde{\sigma%
}_{\pm }$ on the simple roots from which one can easily infer the invariance
of the entire root space. We use the same conventions as for the tables 2
and 3.

\subsection{$A_{8}$-Root spaces based on the class $\Sigma
_{\{1,2,3,4,^{\ell-3}\}}$ and their invariance}

\begin{tabular}{|c|c|c|c|c|c|c|c|c|}
\hline
$\tilde{\sigma}^{(i)}$ & $\alpha_{1}$ & $\alpha_{2}$ & $\alpha_{3}$ & $%
\alpha_{4}$ & $\alpha_{5}$ & $\alpha_{6}$ & $\alpha_{7}$ & $\alpha_{8}$ \\ 
\hline
$\tilde{\sigma}^{(1)}$ & $\mathbf{-1,2}$ & $\mathbf{1,2,3}$ & $\mathbf{-2,3}$
& $2,3,4$ & $5$ & $6$ & $7$ & $8$ \\ \hline
$\tilde{\sigma}^{(1)^{2}}$ & $\mathbf{-3}$ & $\mathbf{-2}$ & $\mathbf{-1}$ & 
$1,2,3,4$ & $5$ & $6$ & $7$ & $8$ \\ \hline
$\tilde{\sigma}^{(1)^{3}}$ & $\mathbf{2,3}$ & $\mathbf{-1,2,3}$ & $\mathbf{%
1,2}$ & $3,4$ & $5$ & $6$ & $7$ & $8$ \\ \hline\hline
$\tilde{\sigma}^{(1)}_{-}$ & $-1$ & $1,2,3$ & $-3$ & $3,4$ & $5$ & $6$ & $7$
& $8$ \\ \hline
$\tilde{\sigma}^{(1)}_{+}$ & $1,2$ & $-2$ & $2,3$ & $4$ & $5$ & $6$ & $7$ & $%
8$ \\ \hline\hline\hline\hline
$\tilde{\sigma}^{(2)}$ & $1,2$ & $\mathbf{3,4}$ & $\mathbf{-2,3,4}$ & $%
\mathbf{2,3}$ & $4,5$ & $6$ & $7$ & $8$ \\ \hline
$\tilde{\sigma}^{(2)^{2}}$ & $1,2,3,4$ & $\mathbf{-4}$ & $\mathbf{-3}$ & $%
\mathbf{-2}$ & $5$ & $6$ & $7$ & $8$ \\ \hline
$\tilde{\sigma}^{(2)^{3}}$ & $1,2,3$ & $\mathbf{-,2,3}$ & $\mathbf{2,3,4}$ & 
$\mathbf{-,3,4}$ & $3,4,5$ & $6$ & $7$ & $8$ \\ \hline\hline
$\tilde{\sigma}^{(2)}_{-}$ & $1$ & $2,3$ & $-3$ & $3,4$ & $5$ & $6$ & $7$ & $%
8$ \\ \hline
$\tilde{\sigma}^{(2)}_{+}$ & $1,2$ & $-2$ & $2,3,4$ & $-4$ & $4,5$ & $6$ & $%
7 $ & $8$ \\ \hline\hline\hline
$\tilde{\sigma}^{(3)}$ & $1$ & $2,3,4$ & $\mathbf{-3,4}$ & $\mathbf{3,4,5}$
& $\mathbf{-4,5}$ & $4,5,6$ & $7$ & $8$ \\ \hline
$\tilde{\sigma}^{(3)^{2}}$ & $1$ & $2,3,4,5$ & $\mathbf{-5}$ & $\mathbf{-4}$
& $\mathbf{-3}$ & $3,4,5,6$ & $7$ & $8$ \\ \hline
$\tilde{\sigma}^{(3)^{3}}$ & $1$ & $2,3$ & $\mathbf{4,5}$ & $\mathbf{-3,4,5}$
& $\mathbf{3,4}$ & $5,6$ & $7$ & $8$ \\ \hline\hline
$\tilde{\sigma}^{(3)}_{-}$ & $1$ & $2,3$ & $-3$ & $3,4,5$ & $-5$ & $5,6$ & $%
7 $ & $8$ \\ \hline
$\tilde{\sigma}^{(3)}_{+}$ & $1$ & $2$ & $3,4$ & $-4$ & $4,5$ & $6$ & $7$ & $%
8$ \\ \hline\hline\hline
$\tilde{\sigma}^{(4)}$ & $1$ & $2$ & $3,4$ & $\mathbf{5,6}$ & $\mathbf{-4,5,6%
}$ & $\mathbf{4,5}$ & $6,7$ & $8$ \\ \hline
$\tilde{\sigma}^{(4)^{2}}$ & $1$ & $2$ & $3,4,5,6$ & $\mathbf{-6}$ & $%
\mathbf{-5}$ & $\mathbf{-4}$ & $4,5,6,7$ & $8$ \\ \hline
$\tilde{\sigma}^{(4)^{3}}$ & $1$ & $2$ & $3,4,5$ & $\mathbf{-4,5}$ & $%
\mathbf{4,5,6}$ & $\mathbf{-5,6}$ & $5,6,7$ & $8$ \\ \hline\hline
$\tilde{\sigma}^{(4)}_{-}$ & $1$ & $2$ & $3$ & $4,5$ & $-5$ & $5,6$ & $7$ & $%
8$ \\ \hline
$\tilde{\sigma}^{(4)}_{+}$ & $1$ & $2$ & $3,4$ & $-4$ & $4,5,6$ & $-6$ & $%
6,7 $ & $8$ \\ \hline\hline\hline
$\tilde{\sigma}^{(5)}$ & $1$ & $2$ & $3$ & $4,5,6$ & $\mathbf{-5,6}$ & $%
\mathbf{5,6,7}$ & $\mathbf{-6,7}$ & $6,7,8$ \\ \hline
$\tilde{\sigma}^{(5)^{2}}$ & $1$ & $2$ & $3$ & $4,5,6,7$ & $\mathbf{-7}$ & $%
\mathbf{-6}$ & $\mathbf{-5}$ & $5,6,7,8$ \\ \hline
$\tilde{\sigma}^{(5)^{3}}$ & $1$ & $2$ & $3$ & $4,5$ & $\mathbf{6,7}$ & $%
\mathbf{-5,6,7}$ & $\mathbf{5,6}$ & $7,8$ \\ \hline\hline
$\tilde{\sigma}^{(5)}_{-}$ & $1$ & $2$ & $3$ & $4,5$ & $-5$ & $5,6,7$ & $-7$
& $7,8$ \\ \hline
$\tilde{\sigma}^{(5)}_{+}$ & $1$ & $2$ & $3$ & $4$ & $5,6$ & $-6$ & $6,7$ & $%
8$ \\ \hline\hline\hline
$\tilde{\sigma}^{(6)}$ & $1$ & $2$ & $3$ & $4$ & $5,6$ & $\mathbf{7,8}$ & $%
\mathbf{-6,7,8}$ & $\mathbf{6,7}$ \\ \hline
$\tilde{\sigma}^{(6)^{2}}$ & $1$ & $2$ & $3$ & $4$ & $5,6,7,8$ & $\mathbf{-8}
$ & $\mathbf{-7}$ & $\mathbf{-6}$ \\ \hline
$\tilde{\sigma}^{(6)^{3}}$ & $1$ & $2$ & $3$ & $4$ & $5,6,7$ & $\mathbf{-6,7}
$ & $\mathbf{6,7,8}$ & $\mathbf{-7,8}$ \\ \hline\hline
$\tilde{\sigma}^{(6)}_{-}$ & $1$ & $2$ & $3$ & $4$ & $5$ & $6,7$ & $-7$ & $%
7,8$ \\ \hline
$\tilde{\sigma}^{(6)}_{+}$ & $1$ & $2$ & $3$ & $4$ & $5,6$ & $-6$ & $6,7,8$
& $-8$ \\ \hline
\end{tabular}
\newpage

\subsection{$A_{8}$-Root spaces based on the class $\Sigma_{\{1,2^{2},3,4,^{%
\ell-4}\}}$ and their invariance}

\begin{tabular}{|c|c|c|c|c|c|c|c|c|}
\hline
$\tilde{\sigma}^{(i,j)}$ & $\alpha_{1}$ & $\alpha_{2}$ & $\alpha_{3}$ & $%
\alpha_{4}$ & $\alpha_{5}$ & $\alpha_{6}$ & $\alpha_{7}$ & $\alpha_{8}$ \\ 
\hline
$\tilde{\sigma}^{(1,1)}$ & $\mathbf{-1,2}$ & $\mathbf{1,2,3}$ & $\mathbf{-2,3%
}$ & $2,3,4,5$ & $-5$ & $5,6$ & $7$ & $8$ \\ \hline
$\tilde{\sigma}^{(1,1)^{2}}$ & $\mathbf{-3}$ & $\mathbf{-2}$ & $\mathbf{-1}$
& $1,2,3,4$ & $5$ & $6$ & $7$ & $8$ \\ \hline
$\tilde{\sigma}^{(1,1)^{3}}$ & $\mathbf{2,3}$ & $\mathbf{-1,2,3}$ & $\mathbf{%
1,2}$ & $3,4,5$ & $-5$ & $5,6$ & $7$ & $8$ \\ \hline\hline
$\tilde{\sigma}^{(1,1)}_{-}$ & $-1$ & $1,2,3$ & $-3$ & $3,4,5$ & $-5$ & $5,6$
& $7$ & $8$ \\ \hline
$\tilde{\sigma}^{(1,1)}_{+}$ & $1,2$ & $-2$ & $2,3$ & $4$ & $5$ & $6$ & $7$
& $8$ \\ \hline\hline\hline
$\tilde{\sigma}^{(2,1)}$ & $1,2$ & $\mathbf{3,4}$ & $\mathbf{-2,3,4}$ & $%
\mathbf{2,3}$ & $4,5,6$ & $-6$ & $6,7$ & $8$ \\ \hline
$\tilde{\sigma}^{(2,1)^{2}}$ & $1,2,3,4$ & $\mathbf{-4}$ & $\mathbf{-3}$ & $%
\mathbf{-2}$ & $2,3,4,5$ & $6$ & $7$ & $8$ \\ \hline
$\tilde{\sigma}^{(2,1)^{3}}$ & $1,2,3$ & $\mathbf{-2,3}$ & $\mathbf{2,3,4}$
& $\mathbf{-3,4}$ & $3,4,5,6$ & $-6$ & $6,7$ & $8$ \\ \hline\hline
$\tilde{\sigma}^{(2,1)}_{-}$ & $1$ & $2,3$ & $-3$ & $3,4$ & $5$ & $6$ & $7$
& $8$ \\ \hline
$\tilde{\sigma}^{(2,1)}_{+}$ & $1,2$ & $-2$ & $2,3,4$ & $-4$ & $4,5,6$ & $-6$
& $6,7$ & $8$ \\ \hline\hline\hline
$\tilde{\sigma}^{(3,1)}$ & $1$ & $2,3,4$ & $\mathbf{-3,4}$ & $\mathbf{3,4,5}$
& $\mathbf{-4,5}$ & $4,5,6,7$ & $-7$ & $7,8$ \\ \hline
$\tilde{\sigma}^{(3,1)^{2}}$ & $1$ & $2,3,4,5$ & $\mathbf{-5}$ & $\mathbf{-4}
$ & $\mathbf{-3}$ & $3,4,5,6$ & $7$ & $8$ \\ \hline
$\tilde{\sigma}^{(3,1)^{3}}$ & $1$ & $2,3$ & $\mathbf{4,5}$ & $\mathbf{-3,4,5%
}$ & $\mathbf{3,4}$ & $5,6,7$ & $-7$ & $7,8$ \\ \hline\hline
$\tilde{\sigma}^{(3,1)}_{-}$ & $1$ & $2,3$ & $-3$ & $3,4,5$ & $-5$ & $5,6,7$
& $-7$ & $7,8$ \\ \hline
$\tilde{\sigma}^{(3,1)}_{+}$ & $1$ & $2$ & $3,4$ & $-4$ & $4,5$ & $6$ & $7$
& $8$ \\ \hline\hline\hline
$\tilde{\sigma}^{(4,1)}$ & $1$ & $2$ & $3,4$ & $\mathbf{5,6}$ & $\mathbf{%
-4,5,6}$ & $\mathbf{4,5}$ & $6,7,8$ & $-8$ \\ \hline
$\tilde{\sigma}^{(4,1)^{2}}$ & $1$ & $2$ & $3,4,5,6$ & $\mathbf{-6}$ & $%
\mathbf{-5}$ & $\mathbf{-4}$ & $4,5,6,7$ & $8$ \\ \hline
$\tilde{\sigma}^{(4,1)^{3}}$ & $1$ & $2$ & $3,4,5$ & $\mathbf{-4,5}$ & $%
\mathbf{4,5,6}$ & $\mathbf{-5,6}$ & $5,6,7,8$ & $-8$ \\ \hline\hline
$\tilde{\sigma}^{(4,1)}_{-}$ & $1$ & $2$ & $3$ & $4,5$ & $-5$ & $5,6$ & $7$
& $8$ \\ \hline
$\tilde{\sigma}^{(4,1)}_{+}$ & $1$ & $2$ & $3,4$ & $-4$ & $4,5,6$ & $-6$ & $%
6,7,8$ & $-8$ \\ \hline
\end{tabular}

\subsection{$A_{8}$-Root spaces based on the class $\Sigma_{\{1,2^{2},3,4,^{%
\ell-4}\}}$ and their invariance}

\begin{tabular}{|c|c|c|c|c|c|c|c|c|}
\hline
$\tilde{\sigma}^{(i,j)}$ & $\alpha_{1}$ & $\alpha_{2}$ & $\alpha_{3}$ & $%
\alpha_{4}$ & $\alpha_{5}$ & $\alpha_{6}$ & $\alpha_{7}$ & $\alpha_{8}$ \\ 
\hline
$\tilde{\sigma}^{(1,2)}$ & $-1$ & $1,2,3,4$ & $\mathbf{-3,4}$ & $\mathbf{%
3,4,5}$ & $\mathbf{-4,5}$ & $4,5,6$ & $7$ & $8$ \\ \hline
$\tilde{\sigma}^{(1,2)^{2}}$ & $1$ & $2,3,4,5$ & $\mathbf{-5}$ & $\mathbf{-4}
$ & $\mathbf{-3}$ & $3,4,5,6$ & $7$ & $8$ \\ \hline
$\tilde{\sigma}^{(1,2)^{3}}$ & $-1$ & $1,2,3$ & $\mathbf{4,5}$ & $\mathbf{%
-3,4,5}$ & $\mathbf{3,4}$ & $5,6$ & $7$ & $8$ \\ \hline\hline
$\tilde{\sigma}^{(1,2)}_{-}$ & $-1$ & $1,2,3$ & $-3$ & $3,4,5$ & $-5$ & $5,6$
& $7$ & $8$ \\ \hline
$\tilde{\sigma}^{(1,2)}_{+}$ & $1$ & $2$ & $3,4$ & $-4$ & $4,5$ & $6$ & $7$
& $8$ \\ \hline\hline\hline
$\tilde{\sigma}^{(2,2)}$ & $1,2$ & $-2$ & $2,3,4$ & $\mathbf{5,6}$ & $%
\mathbf{-4,5,6}$ & $\mathbf{4,5}$ & $6,7$ & $8$ \\ \hline
$\tilde{\sigma}^{(2,2)^{2}}$ & $1$ & $2$ & $3,4,5,6$ & $\mathbf{-6}$ & $%
\mathbf{-5}$ & $\mathbf{-4}$ & $4,5,6,7$ & $8$ \\ \hline
$\tilde{\sigma}^{(2,2)^{3}}$ & $1,2$ & $-2$ & $2,3,4,5$ & $\mathbf{-4,5}$ & $%
\mathbf{4,5,6}$ & $\mathbf{-5,6}$ & $5,6,7$ & $8$ \\ \hline\hline
$\tilde{\sigma}^{(2,2)}_{-}$ & $1$ & $2$ & $3$ & $4,5$ & $-5$ & $5,6$ & $7$
& $8$ \\ \hline
$\tilde{\sigma}^{(2,2)}_{+}$ & $1,2$ & $-2$ & $2,3,4$ & $-4$ & $4,5,6$ & $-6$
& $6,7$ & $8$ \\ \hline
\end{tabular}

\begin{tabular}{|c|c|c|c|c|c|c|c|c|}
\hline
$\tilde{\sigma}^{(3,2)}$ & $1$ & $2,3$ & $-3$ & $3,4,5,6$ & $\mathbf{-5,6}$
& $\mathbf{5,6,7}$ & $\mathbf{-6,7}$ & $6,7,8$ \\ \hline
$\tilde{\sigma}^{(3,2)^{2}}$ & $1$ & $2$ & $3$ & $4,5,6,7$ & $\mathbf{-7}$ & 
$\mathbf{-6}$ & $\mathbf{-5}$ & $5,6,7,8$ \\ \hline
$\tilde{\sigma}^{(3,2)^{3}}$ & $1$ & $2,3$ & $-3$ & $3,4,5$ & $\mathbf{6,7}$
& $\mathbf{-5,6,7}$ & $\mathbf{5,6}$ & $7,8$ \\ \hline\hline
$\tilde{\sigma}^{(3,2)}_{-}$ & $1$ & $2,3$ & $-3$ & $3,4,5$ & $-5$ & $5,6,7$
& $-7$ & $7,8$ \\ \hline
$\tilde{\sigma}^{(3,2)}_{+}$ & $1$ & $2$ & $3$ & $4$ & $5,6$ & $-6$ & $6,7$
& $8$ \\ \hline\hline\hline
$\tilde{\sigma}^{(4,2)}$ & $1$ & $2$ & $3,4$ & $-4$ & $4,5,6$ & $\mathbf{7,8}
$ & $\mathbf{-6,7,8}$ & $\mathbf{6,7}$ \\ \hline
$\tilde{\sigma}^{(4,2)^{2}}$ & $1$ & $2$ & $3$ & $4$ & $5,6,7,8$ & $\mathbf{%
-8}$ & $\mathbf{-7}$ & $\mathbf{-6}$ \\ \hline
$\tilde{\sigma}^{(4,2)^{3}}$ & $1$ & $2$ & $3,4$ & $-4$ & $4,5,6,7$ & $%
\mathbf{-6,7}$ & $\mathbf{6,7,8}$ & $\mathbf{-7,8}$ \\ \hline\hline
$\tilde{\sigma}^{(4,2)}_{-}$ & $1$ & $2$ & $3$ & $4$ & $5$ & $6,7$ & $-7$ & $%
7,8$ \\ \hline
$\tilde{\sigma}^{(4,2)}_{+}$ & $1$ & $2$ & $3,4$ & $-4$ & $4,5,6$ & $-6$ & $%
6,7,8$ & $-8$ \\ \hline
\end{tabular}


\begin{thebibliography}{10}

\bibitem{EW}
E.~Wigner,
\newblock Normal form of antiunitary operators,
\newblock J. Math. Phys. {\bf 1}, 409--413 (1960).

\bibitem{Bender:1998ke}
C.~M. Bender and S.~Boettcher,
\newblock Real Spectra in Non-Hermitian Hamiltonians having PT Symmetry,
\newblock Phys. Rev. Lett. {\bf 80}, 5243--5246 (1998).

\bibitem{Benderrev}
C.~M. Bender,
\newblock Making sense of non-Hermitian Hamiltonians,
\newblock Rept. Prog. Phys. {\bf 70}, 947--1018 (2007).

\bibitem{Milos}
M.~Znojil and M.~Tater,
\newblock Complex Calogero model with real energies,
\newblock J. Phys. {\bf A34}, 1793--1803 (2001).

\bibitem{FZ}
M.~Znojil and A.~Fring,
\newblock PT-symmetric deformations of Calogero models,
\newblock J. Phys. {\bf A41}, 194010(17) (2008).

\bibitem{OP2}
M.~A. Olshanetsky and A.~M. Perelomov,
\newblock Classical integrable finite dimensional systems related to Lie
  algebras,
\newblock Phys. Rept. {\bf 71}, 313--400 (1981).

\bibitem{Wilson}
G.~Wilson,
\newblock The modified Lax and two-dimensional Toda lattice equations
  associated with simple Lie algebras,
\newblock Ergodic Theory and Dynamical Systems {\bf 1}, 361--380 (1981).

\bibitem{DIO}
D.~I. Olive and N.~Turok,
\newblock The symmetries of Dynkin diagrams and the reduction of Toda field
  equations,
\newblock Nucl. Phys. {\bf B215}, 470--494 (1983).

\bibitem{FringSmith}
A.~Fring and M.~Smith,
\newblock Antilinear deformations of Coxeter groups, an application to Calogero
  models,
\newblock J. Phys. {\bf A43}, 325201(28) (2010).

\bibitem{FringSmith2}
A.~Fring and M.~Smith,
\newblock $\cal{PT}$ invariant complex $E_8$ root spaces,
\newblock Int. J. of Theor. Phys. {\bf 50}, 974--981 (2011).

\bibitem{Assis:2009gt}
P.~E.~G. Assis and A.~Fring,
\newblock From real fields to complex Calogero particles,
\newblock J. Phys. {\bf A42}, 425206(14) (2009).

\bibitem{q1}
T.~Oota,
\newblock q-deformed Coxeter element in non-simply laced affine Toda field
  theories,
\newblock Nucl. Phys. {\bf B504}, 738--752 (1997).

\bibitem{q2}
A.~Fring, C.~Korff, and B.~J. Schulz,
\newblock On the universal representation of the scattering matrix of affine
  Toda field theory,
\newblock Nucl. Phys. {\bf B567}, 409--453 (2000).

\bibitem{Ghosh1}
P.~K. Ghosh,
\newblock On the construction of pseudo-hermitian quantum system with a
  pre-determined metric in the Hilbert space,
\newblock J. Phys. {\bf A43}, 125203 (2010).

\bibitem{Ghosh2}
P.~K. Ghosh,
\newblock Deconstructing non-Dirac Hermitian supersymmetric quantum systems,
\newblock J. Phys. {\bf A44}, 215307 (2011).

\bibitem{Kac}
V.~G. Kac,
\newblock Infinite dimensional Lie algebras,
\newblock CUP, Cambridge  (1990).

\bibitem{Bender:2004sa}
C.~M. Bender, D.~C. Brody, and H.~F. Jones,
\newblock Extension of PT-symmetric quantum mechanics to quantum field theory
  with cubic interaction,
\newblock Phys. Rev. {\bf D70}, 025001(19) (2004).

\bibitem{Mist}
A.~Mostafazadeh and A.~Batal,
\newblock Physical Aspects of Pseudo-Hermitian and $PT$-Symmetric Quantum
  Mechanics,
\newblock J. Phys. {\bf A37}, 11645--11680 (2004).

\bibitem{JM}
H.~Jones and J.~Mateo,
\newblock An Equivalent Hermitian Hamiltonian for the non-Hermitian $-x^4$
  Potential,
\newblock Phys. Rev. {\bf D73}, 085002 (2006).

\bibitem{CA}
C.~Figueira~de Morisson~Faria and A.~Fring,
\newblock Time evolution of non-Hermitian Hamiltonian systems,
\newblock J. Phys. {\bf A39}, 9269--9289 (2006).

\bibitem{KBZ}
D.~Krejcirik, H.~Bila, and M.~Znojil,
\newblock Closed formula for the metric in the Hilbert space of a PT-symmetric
  model,
\newblock J. Phys. {\bf A39}, 10143--10153 (2006).

\bibitem{Mostdel}
A.~Mostafazadeh,
\newblock Delta-Function Potential with a Complex Coupling,
\newblock J. Phys. {\bf A39}, 13495--13506 (2006).

\bibitem{MGH}
D.~P. Musumbu, H.~B. Geyer, and W.~D. Heiss,
\newblock Choice of a metric for the non-Hermitian oscillator,
\newblock J. Phys. {\bf A40}, F75--F80 (2007).

\bibitem{PEGA}
P.~E.~G. Assis and A.~Fring,
\newblock Metrics and isospectral partners for the most generic cubic
  PT-symmetric non-Hermitian Hamiltonian,
\newblock J. Phys. {\bf A41}, 244001(18) (2008).

\bibitem{Bender:2008uu}
C.~M. Bender and D.~W. Hook,
\newblock Exact Isospectral Pairs of PT-Symmetric Hamiltonians,
\newblock J. Phys. {\bf A41}, 244005(17) (2008).

\bibitem{PEGA2}
P.~E.~G. Assis and A.~Fring,
\newblock Non-Hermitian Hamiltonians of Lie algebraic type,
\newblock J. Phys. {\bf A42}, 015203(23) (2009).

\bibitem{OlaAnd}
A.~Castro-Alvaredo and A.~Fring,
\newblock A spin chain model with non-Hermitian interaction: the Ising quantum
  spin chain in an imaginary field,
\newblock J. Phys. {\bf A42}(46), 465211(29) (2009).

\bibitem{pseudo1}
M.~Froissart,
\newblock Covariant formalism of a field with indefinite metric,
\newblock Il Nuovo Cimento {\bf 14}, 197--204 (1959).

\bibitem{pseudo2}
E.~C.~G. Sudarshan,
\newblock Quantum Mechanical Systems with Indefinite Metric. I,
\newblock Phys. Rev. {\bf 123}, 2183--2193 (1961).

\bibitem{Mostafazadeh:2001nr}
A.~Mostafazadeh,
\newblock Pseudo-Hermiticity versus PT-Symmetry II: A complete characterization
  of non-Hermitian Hamiltonians with a real spectrum,
\newblock J. Math. Phys. {\bf 43}, 2814--2816 (2002).

\bibitem{Dieu}
J.~Dieudonn{\'{e}},
\newblock Quasi-hermitian operators,
\newblock Proceedings of the International Symposium on Linear Spaces,
  Jerusalem 1960, Pergamon, Oxford , 115--122 (1961).

\bibitem{Urubu}
F.~G. Scholtz, H.~B. Geyer, and F.~Hahne,
\newblock Quasi-Hermitian Operators in Quantum Mechanics and the Variational
  Principle,
\newblock Ann. Phys. {\bf 213}, 74--101 (1992).

\bibitem{Arin}
A.~E. Arinshtein, V.~A. Fateev, and A.~B. Zamolodchikov,
\newblock Quantum s Matrix of the (1+1)-Dimensional Todd Chain,
\newblock Phys. Lett. {\bf B87}, 389--392 (1979).

\bibitem{BCDS}
H.~W. Braden, E.~Corrigan, P.~E. Dorey, and R.~Sasaki,
\newblock Affine Toda field theory and exact S matrices,
\newblock Nucl. Phys. {\bf B338}, 689--746 (1990).

\bibitem{DDV}
C.~Destri and H.~J. de~Vega,
\newblock New exact results in affine Toda field theories: Free energy and wave
  function renormalizations,
\newblock Nucl. Phys. {\bf B358}, 251--294 (1991).

\bibitem{CM1}
P.~Christe and G.~Mussardo,
\newblock Integrable Sytems away from criticality: The Toda field theory and S
  matrix of the tricritical Ising model,
\newblock Nucl. Phys. {\bf B330}, 465--487 (1990).

\bibitem{PD1}
P.~Dorey,
\newblock Root systems and purely elastic S matrices,
\newblock Nucl. Phys. {\bf B358}, 654--676 (1991).

\bibitem{PD2}
P.~Dorey,
\newblock Root systems and purely elastic S matrices. 2,
\newblock Nucl. Phys. {\bf B374}, 741--762 (1992).

\bibitem{Mass2}
A.~Fring, H.~C. Liao, and D.~Olive,
\newblock The mass spectrum and coupling in affine Toda theories,
\newblock Phys. Lett. {\bf B266}, 82--86 (1991).

\bibitem{Mussardo:1992uc}
G.~Mussardo,
\newblock Off critical statistical models: Factorized scattering theories and
  bootstrap program,
\newblock Phys.Rept. {\bf 218}, 215--379 (1992).

\bibitem{Del3}
G.~W. Delius, M.~T. Grisaru, and D.~Zanon,
\newblock Exact S matrices for nonsimply laced affine Toda theories,
\newblock Nucl. Phys. {\bf B382}, 365--408 (1992).

\bibitem{KW}
H.~G. Kausch and G.~M.~T. Watts,
\newblock Duality in quantum Toda theory and W algebras,
\newblock Nucl. Phys. {\bf B386}, 166--192 (1992).

\bibitem{WW}
G.~M.~T. Watts and R.~A. Weston,
\newblock $G_2^{(1)}$ affine Toda field theory: A Numerical test of exact S
  matrix results,
\newblock Phys. Lett. {\bf B289}, 61--66 (1992).

\bibitem{CDS}
E.~Corrigan, P.~E. Dorey, and R.~Sasaki,
\newblock On a generalized bootstrap principle,
\newblock Nucl. Phys. {\bf B408}, 579--599 (1993).

\bibitem{PD}
P.~Dorey,
\newblock A Remark on the coupling dependence in affine Toda field theories,
\newblock Phys. Lett. {\bf B312}, 291--298 (1993).

\bibitem{Khast}
S.~P. Khastgir,
\newblock S-matrices of non-simply laced affine Toda theories by folding,
\newblock Nucl. Phys. {\bf B499}, 650--672 (1997).

\bibitem{Holl}
T.~Hollowood,
\newblock Solitons in affine Toda field theory,
\newblock Nucl. Phys. {\bf B384}, 523--540 (1992).

\end{thebibliography}

\end{document}